\documentclass[11pt,a4paper]{article}
\pdfoutput=1
\usepackage{jheppub} 
\usepackage{amsfonts,amssymb,latexsym,amsmath,mathrsfs}
\usepackage{bm}
\usepackage{slashed,bbold}
\usepackage{hyperref}

\newcommand{\mbD}{\bm{D}}
\newcommand{\mcG}{\mathcal G}
\newcommand{\mbT}{\bm{T}}
\newcommand{\mbJ}{\bm{J}}
\newcommand{\mcB}{\mathcal B}
\newcommand{\mbB}{\bm{B}}
\newcommand{\mbF}{\bm{F}}
\newcommand{\mbA}{\bm{A}}
\newcommand{\mbV}{\bm{V}}
\newcommand{\mcA}{{\mathcal A}}
\newcommand{\mcP}{{\mathcal P}}
\newcommand{\mcJ}{{\mathcal J}}
\newcommand{\mcM}{{\mathcal M}}
\newcommand{\mcV}{{\mathcal V}}

\newcommand{\mcF}{{\mathcal F}}

\newcommand{\x}{{\bf x}}

\usepackage[latin1]{inputenc}

\title{Non-Abelian Anomalous (Super)Fluids in
Thermal Equilibrium from Differential Geometry}

\author[a]{Juan L. Ma\~nes,}
\author[b,c]{Eugenio Meg\'\i as,}
\author[b]{Manuel Valle,}
\author[d]{Miguel \'A. V\'azquez-Mozo}

\affiliation[a]{Departamento de F\'\i sica de la Materia Condensada, 
Universidad del Pa\'is Vasco UPV/EHU, \\
Apartado 644,  48080 Bilbao, Spain}
\affiliation[b]{Departamento de F\'\i sica Te\'orica, 
Universidad del Pa\'is Vasco UPV/EHU, \\
Apartado 644,  48080 Bilbao, Spain}
\affiliation[c]{Departamento de F\'{\i}sica At\'omica, 
Molecular y Nuclear and Instituto Carlos I de F\'{\i}sica Te\'orica y Computacional, 
Universidad de Granada, Avenida de Fuente Nueva s/n,  18071 Granada, Spain
}
\affiliation[d]{Departamento de Física Fundamental, Universidad de Salamanca, 
Plaza de la Merced s/n, \\ E-37008 Salamanca, Spain}

\emailAdd{wmpmapaj@lg.ehu.es}
\emailAdd{eugenio.megias@ehu.eus}
\emailAdd{manuel.valle@ehu.es}
\emailAdd{Miguel.Vazquez-Mozo@cern.ch}

\abstract{
We apply differential geometry methods to the computation of the anomaly-induced hydrodynamic equilibrium partition function.
Implementing the imaginary-time prescription on the Chern-Simons effective action on a stationary background, we 
obtain general closed expressions for both the invariant and anomalous part of the partition function. This is applied to 
the Wess-Zumino-Witten action for Goldstone modes, giving the equilibrium partition function of superfluids. 
In all cases, we also study the anomaly-induced gauge currents and energy-momentum tensor, providing explicit expressions for them.
}


\begin{document}
\maketitle
\flushbottom


\section{Introduction}
\label{sec:intro}

Quantum anomalies play a central role in high energy physics 
(see~\cite{Zumino:1983ew,AlvarezGaume:1985ex,Bertlmann:1996xk,Fujikawa:2004cx,Harvey:2005it} for reviews), being
the cause behind important physical phenomena. They also impose severe constraints on physically viable theories, and
their dual infrared/ultraviolet character can be exploited to 
extract nonperturbative information in a variety of physically interesting situations. 
In recent years it has been realized that anomalies are also relevant 
for hydrodynamics~\cite{Son:2009tf,Neiman:2010zi}. In the presence of anomalous currents coupling to external, nondynamical 
gauge fields, parity is broken and additional tensor structures in the constitutive relations are allowed, associated
with new transport coefficients. Preservation of the second law of thermodynamics leads to a number of identities to be satisfied
by these additional coefficients, ensuring 
the nondissipative character of these anomaly-induced effects. This is of much physical interest, since
anomalous hydrodynamics is 
related to transport phenomena associated with chiral imbalance, 
such as the chiral magnetic and vortical effects~\cite{Sadofyev:2010pr,Kirilin:2012mw}
(more details can be found in the reviews~\cite{Fukushima:2012vr,Zakharov:2012vv,Kharzeev:2015znc}).
There are also important connections to a variety of other condensed matter systems~\cite{Landsteiner:2016led}, giving rise to
experimental signatures \cite{Gooth:2017mbd}.

Alternatively, equilibrium hydrodynamics can be studied without resorting to the entropy
current~\cite{Banerjee:2012iz,Jensen:2012jh,Jensen:2012kj,Jensen:2013kka}. The underlying idea is the construction of an effective action
for the hydrodynamic sources on a generic stationary spacetime, with the appropriate number of derivatives 
and consistent with all relevant symmetries. 
In the case of anomalous hydrodynamics, this effective action functional include additional terms which, upon gauge transformations of the
external sources, render the right value of the anomaly. The 
equilibrium thermal partition function is obtained from this effective action, with the inverse temperature identified 
with the length of the compactified Euclidean time. Correlation functions of currents and other quantities 
are then computed by functional differentiation 
with respect to external sources. 

Since the 1980s, differential geometry has revealed itself 
as a very powerful tool in the analysis of anomalies in quantum field theory~\cite{Zumino:1983ew}. 
The reason for its success lies in the
fact that anomalies are originated in the topological structure of the gauge 
bundle, so they are determined, up to a global normalization, from topological invariant quatities. 
This also extends to the case of systems with spontaneously
symmetry breaking, where the anomaly very much constraints the dynamics of Goldstone bosons \cite{Wess:1971yu,Witten:1983tw,Manes:1984gk}. 
One of the big advantages of
using differential geometry methods in the analysis of anomalies is the possibility of  
constructing effective actions rather straightforwardly in terms of the Chern-Simons forms derived from the appropriate anomaly polynomial.
Using homotopy methods \cite{Zumino:1983ew}, it is possible in many cases to find closed expressions for the Chern-Simons form and its descent quantities. 
Apart from this important fact, a further benefit of the differential geometry approach  
is that it exhausts all perturbative contributions to the 
anomaly, which in the non-Abelian case include not just the triangle, but also the square and pentagon one-loop diagrams. 
These techniques were employed in Refs.~\cite{Jensen:2012kj,Haehl:2013hoa,Jensen:2013kka} to study various aspects of the
physics of anomalous fluids in thermal equilibrium. 

In this article we carry out a systematic 
construction of the equilibrium partition functions for 
fluids with
non-Abelian chiral anomalies in arbitrary even dimension, $D=2n-2$, using differential geometry methods
along the lines
of~\cite{Jensen:2013kka}. Our aim is to provide explicit expressions that could be easily applied to the 
study of nondissipative anomalous fluids. To this end, we give a general prescription
to obtain the anomaly-induced part of the partition function by performing a dimensional reduction on the time circle of the
Chern-Simons effective action defined on a $(2n-1)$-dimensional stationary spacetime. With this we immediately show that the
partition function splits into a {\em local} anomalous piece and a {\em nonlocal} invariant part, as pointed out in~\cite{Jensen:2013kka}.
Moreover, we provide operative expressions for both contributions to the partition function that can be used in the analysis of generic theories.
Their use is illustrated with various examples in four dimensions~($n=3$).

One of the main targets of our analysis is the study of hydrodynamics in the presence of spontaneous symmetry 
breaking~\cite{Lin:2011mr,Nair:2011mk,Bhattacharyya:2012xi}, extending the analysis of \cite{Jensen:2013kka} to this setup. The
equilibrium partition function for these systems is built from the dimensional reduction of the Wess-Zumino action describing the dynamics of 
the Goldstone modes on a generic stationary background. Unlike the unbroken case, 
here we see that the effective action only depends on the local anomalous
part of the action. 
The general expressions obtained this way might be of relevance for non-Abelian anomalous 
superfluids~\cite{Lublinsky:2009wr,Neiman:2011mj,Lin:2011aa,Hoyos:2014nua,Jain:2016rlz}. 

In all cases, we provide explicit expression for both the gauge currents and the energy-momentum tensor.
When the symmetry remains unbroken, the covariant current is computed from the variation of the nonlocal invariant part of the action, whereas
the consistent current is obtained from the anomalous local piece \cite{Jensen:2013kka}. In the presence of 
spontaneous symmetry breaking, on the other hand, we find that both the covariant and the consistent currents can be written in terms
of the Bardeen-Zumino current interpolating between the two. We thus compute the currents on a stationary background from the 
dimensional reduction of the Bardeen-Zumino current. Moreover, an evaluation of the leading anomaly-induced energy-momentum tensor gives a
vanishing result, due to the cancellation between the contributions of gauge fields and Goldstone modes.

The power of the techniques presented here lie on their systematic and wide applicability, 
providing a fast and efficient way of computing the equilibrium partition function and
the transport coefficients in generic models. We will illustrate this point with 
some sample applications. A systematic analysis of particular models, including full details, will be presented  
in a future work~\cite{in_progress}.

The remainder of the article is organized as follows. In Section~\ref{sec:anom} we offer a quick review of the basic aspects of the
differential geometry approach to the construction of anomalous effective actions. 
We pay special attention to the case of two fermions with different chirality
coupled to independent external gauge fields and, using the generalized transgression formula~\cite{Manes:1985df}, 
we compute the effective action giving the anomaly in the Bardeen form. The transgression formula is further used in Section \ref{sec:CS} to 
define the currents associated with the Chern-Simons effective action. In Section \ref{sec:anomalousPF}, we construct the fluid equilibrium 
partition function on a generic stationary background by dimensional reduction of the Chern-Simons effective action on the time cycle. In this
way, we show that the anomalous part of the dimensionally-reduced partition function becomes local. Our results are then applied to the 
four-dimensional Bardeen anomaly and, as a particular case, we obtain the partition function of a two-flavor hadronic fluid.  

In Section \ref{sec:currents_em_tensor} we study the gauge currents and the energy-momentum tensor derived from the 
anomaly-induced partition function on a general
stationary background. We obtain the Wess-Zumino term relating the consistent and the covariant currents. 
The expressions found in this section are illustrated with the results for two-flavor QCD coupled to an
external electromagnetic field on a nontrivial stationary background.
Section~\ref{sec:WZW} is devoted to the study of anomalous fluids in theories with spontaneous symmetry breaking. We construct the
partition function at leading order in derivatives, starting from the Wess-Zumino-Witten effective action of Goldstone bosons on a stationary 
metric and implementing the imaginary time prescription.
After discussing the issue of the currents in this setup, we compute the partition function in four-dimensions for a two-flavor 
hadronic superfluid where the global flavor group is broken down to its vector subgroup. We also compute the associated currents
and anomaly-induced energy-momentum tensor. Finally, in Section \ref{sec:discussion}
we summarize our results and discuss future lines of work.

To make the article self-contained, a short review of the generalized transgression formula is presented in Appendix~\ref{app:transgression}, 
whereas some explicit expressions are deferred to Appendix \ref{app:formulae}. 
Finally, in Appendix \ref{app:traces} we have summarized some relevant trace identities for the group U(2). 


\section{Anomalies and differential geometry}
\label{sec:anom}

Differential geometry is a very powerful tool in the study of quantum field theory anomalies. One of its advantages is its
power in providing very general prescriptions to construct quantum effective actions.
In this section we are going to review basic aspects of the differential geometry approach to quantum anomalies to 
be applied in the rest of the paper. More details can be found in the 
reviews~\cite{Zumino:1983ew,AlvarezGaume:1985ex,Bertlmann:1996xk,Fujikawa:2004cx,Harvey:2005it}. 

\subsection{Chiral anomalies and effective actions}
\label{sec:LRsymm_anom}

We begin by studying the theory of a chiral fermion coupled to an external gauge field 
$\mcA_{\mu}\equiv  \mcA_{\mu}^{a}t_{a}$ described by the Lagrangian 
\begin{align}
\mathscr{L}_{\rm YM}=i\overline{\psi}\gamma^{\mu}\Big(\partial_{\mu}-it_{a}\mcA^{a}_{\mu}\Big)\psi,
\label{eq:YMLagrangian}
\end{align} 
where $t_{a}=t_{a}^\dagger$ are the Hermitian generators of the Lie algebra $\mathfrak{g}=\mbox{Lie}(\mathcal{G})$ satisfying the commutation relations
$ [t_{a},t_{b}]=if_{abc}t_{c}$. Here and in the following, the gauge coupling constant $g$ is absorbed into the gauge field $\mcA_{\mu}^{a}$. 
It is convenient to introduce  the anti-Hermitian, Lie algebra valued one-form\footnote{In this paper we systematically 
omit the wedge symbol $\wedge$ in the exterior product of differential forms. Otherwise, we
follow the conventions of Ref.~\cite{Nakahara:2003nw}.}
\begin{align}
\mcA=-i\mcA^{a}_{\mu}t_{a} dx^{\mu}\equiv -i\mcA_\mu dx^{\mu},
\label{eq:A_def}
\end{align} 
while the associated field strength two-form is defined by
\begin{align}
\mcF=d\mcA+\mcA^2\equiv -{i\over 2}\mcF_{\mu\nu}dx^{\mu}dx^{\nu},
\label{eq:F_two-form_def}
\end{align}
with components
\begin{align}
\mcF_{\mu\nu}=\partial_{\mu}\mcA_{\nu}-\partial_{\nu}\mcA_{\mu}-i[\mcA_{\mu},\mcA_{\nu}].
\end{align}
The field strength two-form satisfies the Bianchi identity
\begin{align}
d\mcF=[\mcF,\mcA].
\label{eq:bianchi_id}
\end{align}

Finite gauge transformations are implemented by the Lie group elements
\begin{align}
g=e^{-it_{a}u^{a}}\equiv e^{u},
\end{align}
where $u(x)$ is a Lie algebra valued function depending on the spacetime point. The gauge field one-form $\mcA$ transforms as a connection
\begin{align}
\mcA_{g}=g^{-1}\mcA g+g^{-1}dg,
\end{align}
whereas the field strength transforms covariantly as an adjoint field
\begin{align}
\mcF_{g}=g^{-1}\mcF g. 
\end{align}
The corresponding infinitesimal transformations are obtained by expanding the previous expressions to leading order in the
gauge function $u(x)$, to give 
\begin{align}
\delta_{u}\mcA&=du+[\mcA,u]\equiv Du, \nonumber \\[0.2cm]
\delta_{u}\mcF&=[\mcF,u],
\label{eq:AFgauge_trans}
\end{align}
where we have introduced the adjoint covariant derivative $Du$. 

To study gauge anomalies it is convenient to work with the fermion effective action functional obtained by integrating out the 
fermion field
\begin{align}
e^{i\Gamma[\mcA]}\equiv \int \mathscr{D}\overline{\psi}\mathscr{D}\psi\, e^{iS_{\rm YM}[\mcA,\psi,\overline{\psi}]}.
\end{align}
The gauge anomaly is then signalled by the transformation of this effective action under gauge transformation. Indeed, under a general
shift $\mcA_{\mu}^{a}\rightarrow \mcA_{\mu}^{a}+\delta \mcA_{\mu}^{a}$, the consistent current is defined from the first order variation of
$\Gamma[\mcA]$ by
\begin{align}
\delta \Gamma[\mcA]=\int d^{D}x\,\delta \mcA_\mu^a(x)J_a^\mu(x),
\label{eq:jcons}
\end{align}
whereas the anomaly is given by the failure of the effective functional to be invariant under gauge transformations
\begin{align}
\delta_u \Gamma[\mcA]=-\int d^{D}x\,u^{a}(x)G_{a}[\mcA(x)]_{\rm cons},
\label{eq:noninv}
\end{align}
where $G_{a}[\mcA(x)]_{\rm cons}$ is the consistent anomaly.
Particularizing now Eq. \eqref{eq:jcons} to gauge transformations, $\delta\mcA_\mu^a=(D_{\mu}u)^{a}$, we arrive at the 
anomalous (non)conservation law for the consistent gauge current
\begin{align}
D_\mu J^\mu_a(x)_{\rm cons}=G_{a}[\mcA(x)]_{\rm cons}.
\label{eq:consanom}
\end{align}

This form of the anomaly as well as the current defined in Eq. \eqref{eq:jcons} are called \emph{consistent} because the
anomaly satisfies the Wess-Zumino consistency
condition implied by the closure of the commutator of two infinitesimal gauge transformations 
\begin{align}
\delta_{u}\delta_{v}-\delta_{v}\delta_{u}=\delta_{[u,v]}.
\label{eq:WZ_const_general}
\end{align}
Indeed, from the definition of the consistent anomaly in terms of the gauge variation of the effective functional 
given in \eqref{eq:noninv}, we automatically arrive at the Wess-Zumino consistency conditions~\cite{Wess:1971yu} by applying 
Eq. \eqref{eq:WZ_const_general} to the effective action
\begin{align}
\int d^{D}x\,v^{a}\delta_{u}G_{a}[\mcA]
-\int d^{D}x\,u^{a}\delta_{v}G_{a}[\mcA]=
\int d^{D}x\,[u,v]^{a}G_{a}[\mcA].
\label{eq:WZconsistency_cond_final}
\end{align}

In fact, the gauge variation of any functional of the gauge field will automatically satisfy the Wess-Zumino consistency
conditions. This applies to the consistent anomaly, which being obtained as the gauge variation of $\Gamma[\mcA]$
fulfills them naturally.
However, not every solution to the consistency condition \eqref{eq:WZconsistency_cond_final}
gives an anomaly. For a theory to be anomalous, the anomaly should be given as the gauge variation of a
\emph{nonlocal} functional of the gauge field. Otherwise the violation in the conservation of the gauge current can be cancelled by 
adding a local counterterm to the action (i.e., by imposing an appropriate renormalization condition).

It is  well known that the non-Abelian gauge anomaly $G_{a}[\mcA]$ for a chiral fermion can be obtained from the corresponding anomaly polynomial by using the Stora-Zumino descent equations~\cite{Zumino:1983ew}. For a right-handed fermion in $D=2n-2$ spacetime dimensions,
the corresponding anomaly polynomial is given by
\begin{align}
\mathcal{P}_{n}(\mcF)=c_n{\rm Tr\,}\mcF^{n},
\end{align}
where the correct normalization $c_{n}$ is found by applying the Atiyah-Singer index theorem to certain Dirac operator
in $2n=D+2$ dimensions~\cite{AlvarezGaume:1983cs}
\begin{align}
c_n=\frac{1}{n!}\frac{i^n}{(2\pi)^{n-1}}.
\label{eq:c_n_constant}
\end{align}

Using the Bianchi identity \eqref{eq:bianchi_id} it is easy to show that the anomaly polynomial is a closed form
\begin{align}
d\,{\rm Tr\,}\mcF^{n}=0,
\label{eq:dFn=0}
\end{align}
whereas the gauge transformation \eqref{eq:AFgauge_trans} implies that it is gauge invariant
\begin{align}
\delta_{u}{\rm Tr\,}\mcF^{n}=0.
\label{eq:deltaFn=0}
\end{align}
Invoking the Poincar\'e lemma, we conclude from the identity \eqref{eq:dFn=0} that ${\rm Tr\,}\mcF^{n}$ is locally exact. Thus, we write
\begin{align}
{\rm Tr\,}\mcF^{n}=d\omega_{2n-1}^{0}(\mcA),
\label{eq:def_omega0_2n-1_single_field}
\end{align}
where $\omega_{2n-1}^{0}(\mcA)$ is the Chern-Simons form. Equation \eqref{eq:deltaFn=0}, on the other hand, implies also that the gauge variation 
of $\omega_{2n-1}^{0}$ is, locally, a total differential
\begin{align}
\delta_{u}\omega^{0}_{2n-1}(\mcA)
=d \omega^{1}_{2n-2}(u,\mcA).
\label{eq:descent}
\end{align}
This identity is the key ingredient to identify the consistent anomaly. The fermion effective action is constructed as the integral of the
Chern-Simons form 
\begin{align}
\Gamma[\mcA]_{\rm CS}=c_{n}\int\limits_{\mathcal{M}_{2n-1}} \omega^{0}_{2n-1}(\mcA),
\label{eq:CS_eff_action}
\end{align}
where $\mathcal{M}_{2n-1}$ is a $(2n-1)$-dimensional manifold whose boundary $\partial\mathcal{M}_{2n-1}$
corresponds to the $(2n-2)$-dimensional spacetime manifold. 
Its gauge variation
is computed using \eqref{eq:descent} and the Stokes theorem to give
\begin{align}
\delta_{u}\Gamma[\mcA]_{\rm CS}&=c_{n}\int\limits_{\mathcal{M}_{2n-1}} \delta_{u}\omega^{0}_{2n-1}(\mcA)
=c_{n}\int\limits_{\mathcal{M}_{2n-1}}d\omega^{1}_{2n-2}(u,\mcA) \nonumber \\[0.2cm]
&=c_{n}\int\limits_{\partial\mathcal{M}_{2n-1}}\omega^{1}_{2n-2}(u,\mcA).
\end{align}
Comparing with Eq. \eqref{eq:noninv}, we find that 
$\omega^{1}_{2n-2}(u,\mcA)$ yields the properly normalized 
anomaly for a right-handed fermion as
\begin{align}
-c_n\,\omega^{1}_{2n-2}(u,\mcA)=u^aG_a[\mcA]\equiv {\rm Tr\,}\Big(uG[\mcA(x)]_{\rm cons}\Big).
\label{eq:anom}
\end{align}
Moreover, since it results from a gauge variation of a functional, it automatically satisfies the Wess-Zumino consistency condition, hence
the subscript on the right-hand side of this equation. The crucial element in this analysis is that, being an integral over a 
higher-dimensional manifold, the effective action \eqref{eq:CS_eff_action} is \emph{nonlocal} in dimension $D=2n-2$ 
and provides a nontrivial solution
to the consistency condition.   
Incidentally, Eq. \eqref{eq:anom} is also valid for a left-handed fermion with the replacement $c_n\rightarrow -c_n$.

From the previous discussion, we see that the computation of the gauge anomaly boils down to the evaluation of the Chern-Simons form.
This object can be expressed in a closed form using the homotopy formula
\begin{align}
\omega^{0}_{2n-1}(\mcA)=n\int_{0}^{1}dt\,{\rm Tr\,}\Big(\mcA\mcF_{t}^{n-1}\Big),
\label{eq:omega0_2n-1_one_field}
\end{align} 
where $\mcF_t=d\mcA_t+\mcA_t^2$ and $\mcA_t=t \mcA$ is a family of gauge connections continuously interporlating between $\mcA_0=0$ and 
$\mcA_1=\mcA$. The same interpolation can be used to obtain the expression~\cite{Zumino:1983ew}
\begin{align}
\omega^{1}_{2n-2}(u,\mcA)=n\int_{0}^{1}dt(1\!-\!t)\,{\rm Tr\,}\Big[ud\Big(\mcA\mcF_{t}^{n-2}+\mcF_{t}\mcA\mcF_{t}^{n-3}
+\ldots+\mcF_{t}^{n-2}\mcA\Big)\Big].
\end{align} 
In $D=4$ spacetime dimensions ($n=3$), the above formulae yield the five-dimensional Chern-Simons form
\begin{align}
\omega^{0}_{5}(\mcA)={\rm Tr\,}\left(\mcA\mcF^{2}-{1\over 2}\mcA^{3}\mcF+{1\over 10}\mcA^{5}\right),
\end{align} 
and the consistent anomaly 
\begin{align}
{\rm Tr\,}\Big(uG[\mcA(x)]_{\rm cons}\Big)={i\over 24\pi^{2}}\omega^{1}_{4}(u,\mcA)
={i\over 24\pi^{2}}{\rm Tr\,}\left[ud\left(\mcA\mcF-{1\over 2}\mcA^{3}\right)\right].
\end{align} 

\subsection{The Bardeen form of the anomaly}

In this paper we are particularly interested in a more general theory with chiral fermions\
$\psi_{L}$, $\psi_{R}$ respectively coupled  to 
two external gauge fields that we denote by $\mcA_{L}$ and $\mcA_{R}$
\begin{align}
\mathscr{L}_{\rm YM}=i\overline{\psi}_{L}\gamma^{\mu}\Big(\partial_{\mu}-it_{a}\mcA^{a}_{L\mu}\Big)\psi_{L}
+i\overline{\psi}_{R}\gamma^{\mu}\Big(\partial_{\mu}-it_{a}\mcA^{a}_{R\mu}\Big)\psi_{R}.
\label{eq:lagrangian_chiral}
\end{align}
Taking into account that left-handed fermions contribute with a relative minus sign, the anomaly polynomial is given by
\begin{align}
\mathcal{P}_{n}(\mcF_R,\mcF_L)=c_n\Bigl({\rm Tr\,}\mcF_R^{n}-{\rm Tr\,}\mcF_L^{n}\Bigr), 
\label{eq:invpol}
\end{align}
where the normalization constant $c_{n}$ is given in Eq. \eqref{eq:c_n_constant} and $\mcF_{R,L}$ are defined in terms of
$\mcA_{R,L}$ by the usual relation \eqref{eq:F_two-form_def}.

Since the field strengths for the left and right gauge fields 
satisfy the Bianchi identity \eqref{eq:bianchi_id}, the anomaly polynomial is closed. Thus, the Chern-Simons form must be locally exact
\begin{align}
{\rm Tr\,}\mcF_R^{n}-{\rm Tr\,}\mcF_L^{n}=d\omega_{2n-1}^{0}(\mcA_R,\mcA_L).
\label{eq:anom_polynomial_omega02n-1}
\end{align}
Using the generalized transgression formula derived in~\cite{Manes:1985df}
(see Appendix \ref{app:transgression}
 for a quick review of the technique), the Chern-Simons form in the right-hand side of this equation can be written as
[cf. Eq. \eqref{eq:omega0_2n-1_one_field}]
\begin{align}
\omega^{0}_{2n-1}(\mathcal{A}_{R,L},\mathcal{F}_{R,L})
=n\int_{0}^{1}dt\,{\rm Tr\,}\Big(\dot{\mathcal{A}}_{t}\mathcal{F}_{t}^{n-1}\Big),
\label{eq:omega0_2n-1_RL_general}
\end{align}
where $\mathcal{F}_{t}$ is the field strength associated with a one-parameter family of connections $\mathcal{A}_{t}$
interpolating between $\mathcal{A}_{0}=\mathcal{A}_{L}$ and $\mathcal{A}_{1}=\mathcal{A}_{R}$. The dot represents
the derivative with respect to $t$. 

At this point, we have different choices for the family of connections. One possibility is to take
\begin{align}
\mathcal{A}_{t}=(1-2t)\vartheta\left({1\over 2}-t\right)\mathcal{A}_{L}
+(2t-1)\vartheta\left(t-{1\over 2}\right)\mathcal{A}_{R},
\end{align}
with $\vartheta(x)$ the Heaviside step function. Computing the integral in \eqref{eq:omega0_2n-1_RL_general}, we arrive at
the result
\begin{align}
\omega_{2n-1}^{0}(\mcA_R,\mcA_L)=\omega_{2n-1}^{0}(\mcA_R)-\omega_{2n-1}^{0}(\mcA_L),
\label{eq:CS_Sym}
\end{align}
from which we construct the left-right symmetric form of the anomaly. Our original Lagrangian 
\eqref{eq:lagrangian_chiral} is invariant under $\mathcal{G}_{R}\times\mathcal{G}_{L}$ gauge 
transformations generated by $(u_R,u_L)$ and acting independently on $(\mcA_R,\mcA_L)$.
The solution \eqref{eq:CS_Sym}, however, leads to an effective action $\Gamma[\mcA_{R},\mcA_{L}]_{\rm CS}$ which
does not remain invariant under vector gauge transformations, defined as those for which $u_{R}=u_{L}\equiv u_{V}$. 
As it is well known from the general theory of anomalies, we can impose the conservation of the vector current
as a renomalization condition. In the language of differential geometry, this ambiguity amounts to adding a exact 
differential on the right-hand
side of \eqref{eq:CS_Sym}
\begin{align}
\widetilde{\omega}_{2n-1}^{0}(\mcA_R,\mcA_L)=\omega_{2n-1}^{0}(\mcA_R)-\omega_{2n-1}^{0}(\mcA_L)
+dS_{2n-2}(\mcA_{R},\mcA_{L}),
\end{align}
which does not modify the identity \eqref{eq:anom_polynomial_omega02n-1}. The explicit expression of the Bardeen counterterm
$S_{2n-2}$ leading to the preservation of vector gauge transformations can be found in Appendix \ref{app:formulae}.

In the present formalism, this choice of $S_{2n-2}$ amounts to selecting a particular   
interpolating curve in space of gauge connections, which turns out to be
\begin{align}
\mathcal{A}_{t}=(1-t)\mathcal{A}_{L}+t\mathcal{A}_{R}.
\end{align}
Thus, the corresponding Chern-Simons form is
\begin{align}
\widetilde\omega_{2n-1}^{0}(\mcA_R,\mcA_L)=n\int_{0}^{1}dt\,{\rm Tr\,}\Big[(\mcA_R-\mcA_L)\mcF_{t}^{n-1}\Big],
\label{om5_vector}
\end{align}
where in this case $\mcF_{t}$ is explicitly given by
\begin{align}
\mcF_{t}=(1-t)\mcF_{L}+t\mcF_{R}+t(t-1)(\mcA_{R}-\mcA_{L})^{2}.
\end{align} 
Since both $\mcA_R-\mcA_L$ and $\mcF_t$ transform covariantly as adjoint fields under 
vector gauge transformations, the effective
action constructed from \eqref{om5_vector} is automatically invariant under this subgroup. The anomaly can be 
obtained now by applying a generic gauge transformation to $\widetilde\omega_{2n-1}^{0}(\mcA_R,\mcA_L)$
\begin{align}
\delta_{u_{R},u_{L}}\widetilde{\omega}^{0}_{2n-1}(\mcA_R,\mcA_L)=d\widetilde\omega_{2n-2}^{1}(u_{R},u_{L},\mcA_R,\mcA_L).
\end{align} 
Using again the generalized transgression formula, it is possible to write $\widetilde{\omega}^{1}_{2n-2}$ in a compact 
expression  
\begin{align}
\widetilde{\omega}^{1}_{2n-2}(u_{L,R},\mcA_{L,R},\mcF_{L,R})&=n\int_{0}^{1}dt\,{\rm Tr\,}\Bigg\{
(u_{R}-u_{L})\Bigg[\mcF_{t}^{n-1} 
\label{om4_vector}
\\[0.2cm]
&+t(t-1)\sum_{k=0}^{n-2}\Big\{\mcA_{R}-\mcA_{L},
\mcF_{t}^{n-k-2}(\mcA_{R}-\mcA_{L})\mcF_{t}^{k}\Big\}
\Bigg]\Bigg\},
\nonumber
\end{align}
which gives the anomaly in the Bardeen or conserved vector form~\cite{Bardeen:1969md}. The dependence on the gauge functions
makes it manifest that the anomaly vanishes for vector gauge transformations $u_{R}=u_{L}$. 

Later in the paper we will consider the Bardeen form of the anomaly written in terms of vector and axial-vector gauge fields $(\mcV,\mcA)$, defined in terms of $(\mcA_{R},\mcA_{L})$ by
\begin{align}
\mathcal{A}_R &\equiv \mathcal{V} + \mathcal{A},   \nonumber \\[0.2cm]
\mathcal{A}_L &\equiv \mathcal{V} -\mathcal{A},
\end{align}
while the corresponding field strengths are related by
\begin{align}
\mathcal{F}_R &\equiv d \mathcal{A}_R + \mathcal{A}_R^2 = 
 \mathcal{F}_V +\mathcal{F}_A, \nonumber  \\[0.2cm] 
 \mathcal{F}_L &\equiv d \mathcal{A}_L + \mathcal{A}_L^2 =
\mathcal{F}_V -\mathcal{F}_A.
\end{align}
Here $\mathcal{F}_{V}$ and $\mathcal{F}_{A}$ are respectively given by
\begin{align}
\mathcal{F}_V &= -{i\over 2} \mathcal{V}_{\mu \nu} dx^\mu dx^\nu = d \mathcal{V} + \mcV^2 +   \mcA^2, \nonumber \\[0.2cm]
\mathcal{F}_A &= -{i\over 2} \mathcal{A}_{\mu \nu} dx^\mu dx^\nu = d \mathcal{A} + \mcA \mcV +   \mcV \mcA , 
\end{align}
where the components of the corresponding field strengths take the form
\begin{align}
\mcV_{\mu \nu} &=  \partial_\mu \mcV_\nu - \partial_\nu \mcV_\mu - i[\mcV_\mu, \mcV_\nu] - i[\mcA_\mu, \mcA_\nu],  
\nonumber \\[0.2cm]
\mcA_{\mu \nu} &=  \partial_\mu \mcA_\nu - \partial_\nu \mcA_\mu - i[\mcV_\mu, \mcA_\nu] - i[\mcA_\mu, \mcV_\nu].
\end{align}
 
In order to write the original Lagrangian \eqref{eq:lagrangian_chiral} in terms of the vector and axial-vector gauge fields,
we combine left- and right-handed Weyl spinors $\psi_{L}, \psi_{R}$ into a single Dirac field $\psi$ as
\begin{align}
\psi_R&=\frac{1}{2}(1+\gamma_5)\psi, \nonumber \\[0.2cm]  
\psi_L&=\frac{1}{2}(1-\gamma_5)\psi,
\end{align}
so the Lagrangian becomes
\begin{align}
\mathscr{L}_{\rm YM}=i\overline{\psi}\gamma^{\mu}\Big(\partial_{\mu}-it_{a}\mcV^{a}_{\mu}-i\gamma_5t_{a}\mcA^{a}_{\mu}\Big)\psi .
\label{eq:YMLagrangianVA}
\end{align} 
In terms of these fields, the Chern-Simons form in \eqref{om5_vector} is recast as
\begin{align}
\widetilde\omega_{2n-1}^{0}(\mcV,\mcA)=2n\int_{0}^{1}dt\,{\rm Tr\,}\Big(\mcA\mcF_{t}^{n-1}\Big),
\label{eq:omega0_2n-1VA}
\end{align}
with
\begin{align}
\mcF_{t}=\mcF_{V}+(2t-1)\mcF_{A}+4t(t-1)\mcA^{2}.
\end{align}
Similarly, Eq. \eqref{om4_vector} giving the Bardeen anomaly can be rewritten 
in terms of axial and vector gauge fields in the following closed form
\begin{align}
\widetilde{\omega}^{1}_{2n-2}(u_{A},\mcV,\mcA)&=2n\int_{0}^{1}dt\,{\rm Tr\,}\Bigg\{
u_{A}\Bigg[\mathcal{F}_{t}^{n-1}
+4t(t-1)\sum_{k=0}^{n-2}\Big\{\mcA,\mcF_{t}^{n-k-2}\mcA\mcF_{t}^{k}\Big\}\Bigg]
\Bigg\},
\label{om4_vector_VA}
\end{align}
where we have defined
\begin{align}
u_R&=u_V+u_A, \nonumber \\[0.2cm] 
u_L&=u_V-u_A.
\end{align}
These functions generate infinitesimal vector and axial-vector gauge transformations given by
\begin{equation}
\label{eq:transf}
\begin{split}
\delta_{V,A} \mathcal{V} & = du_V + [  \mathcal{V},u_V ] + [  \mathcal{A},u_A],   \\[0.2cm]
\delta_{V,A} \mathcal{A} & = du_A + [  \mathcal{A},u_V ] + [  \mathcal{V}, u_A]  .
\end{split}
 \end{equation}
In particular, $\mathcal{A}$, $\mathcal{F}_{V}$, and $\mathcal{F}_{A}$ transform as adjoint fields under
vector gauge transformations ($u_{A}=0$). Thus, the Chern-Simons form \eqref{eq:omega0_2n-1VA} remains invariant under
them, while the anomaly \eqref{om4_vector_VA} only depends on $u_{A}$. 

For later applications, we particularize Eqs. \eqref{eq:omega0_2n-1VA} and \eqref{om4_vector_VA} 
to the four-dimensional case $(n=3)$. 
In terms of the fields $(\mcV,\mcA)$, the relevant anomaly polynomial is given by
\begin{align}
\mcP_{3}(\mcF_V,\mcF_A)=-{i\over 12\pi^{2}}{\rm Tr\,}\Big(\mcF_A^3+3\mcF_A\mcF_V^2\Big).
\label{eq:invpol3}
\end{align}
After a few manipulations, the Chern-Simons form preserving the vector Ward identity is obtained as
\begin{align}
\widetilde{\omega}^{0}_{5}(\mathcal{A},\mathcal{F}_{V},
\mathcal{F}_{A})&=6{\rm Tr\,}\left(\mathcal{A}\mathcal{F}_{V}^{2}
+{1\over 3}\mathcal{A}\mathcal{F}_{A}^{2}-{4\over 3}\mathcal{A}^{3}\mathcal{F}_{V}
+{8\over 15}\mathcal{A}^{5}\right),
\label{eq:omega05AV_basis}
\end{align} 
while the celebrated Bardeen anomaly~\cite{Bardeen:1969md} is obtained from
\begin{align}
\widetilde{\omega}^{1}_{4}(u_{A},\mathcal{A},\mathcal{F}_{V},\mathcal{F}_{A})&=
6{\rm Tr\,}\left\{u_{A}\left[\mathcal{F}_{V}^{2}+{1\over 3}\mathcal{F}_{A}^{2}
\right.\right. \nonumber \\[0.2cm]
&-{4\over 3}\Big(\mathcal{A}^{2}\mathcal{F}_{V}+\mathcal{A}\mathcal{F}_{V}\mathcal{A}
+\mathcal{F}_{V}\mathcal{A}^{2}\Big)+\left.\left.{8\over 3}\mathcal{A}^{4}\right]\right\}.
\label{eq:omega14AV_basis}
\end{align}
As already pointed out, the absence of $u_V$ in this expression makes explicit the conservation of
the consistent vector current $J_{V}^{\mu}(x)_{\rm cons}$. 

In the case of a two flavor hadronic fluid, the chiral group is U(2)$_{L}\times$U(2)$_{R}$ and traces can be computed
using the identities given in Appendix \ref{app:traces}. For the Chern-Simons form~\eqref{eq:omega05AV_basis}, we have
\begin{align}
\widetilde{\omega}^{0}_{5}(\mathcal{A},\mathcal{F}_{V},
\mathcal{F}_{A})&={3\over 2}({\rm Tr\,}\mathcal{F}_{V})^{2}({\rm Tr\,}\mathcal{A})+
3({\rm Tr\,}\widehat{\mathcal{F}}_{V}^{2})({\rm Tr\,}\mathcal{A})+
6{\rm Tr\,}(\widehat{\mathcal{A}}\widehat{\mathcal{F}}_{V})({\rm Tr\,}\mathcal{F}_{V}) \nonumber\\[0.2cm]
&+{1\over 2}({\rm Tr\,}\mathcal{F}_{A})^{2}({\rm Tr\,}\mathcal{A})+
({\rm Tr\,}\widehat{\mathcal{F}}_{A}^{2})({\rm Tr\,}\mathcal{A})+
2{\rm Tr\,}(\widehat{\mathcal{A}}\widehat{\mathcal{F}}_{A})({\rm Tr\,}\mathcal{F}_{A}) \\[0.2cm]
&-4{\rm Tr\,}(\widehat{\mathcal{A}}^{\,2}\widehat{\mathcal{F}}_{V})({\rm Tr\,}\mathcal{A})
-4({\rm Tr\,}\widehat{\mathcal{A}}^{\,3})({\rm Tr\,}\mathcal{F}_{V}),
\nonumber
\end{align}
where the hat indicates the projection onto the SU(2) factors [see Eq. \eqref{eq:hat_definition}]. We notice here
the absence of terms containing one single trace over SU(2) indices. The reason is that SU(2) is a safe group, so anomalies can
only appear from one-loop diagrams containing at least one vertex coupling to a U(1) factor.  
We can assume that the axial-vector external field $\mcA$, as well as the associated field strength $\mcF_{A}$, 
lies on the SU(2)$_{A}$ factor. This means that ${\rm Tr\,}\mcA={\rm Tr\,}\mcF_{A}=0$ and the Chern-Simons 
form simplifies to
\begin{align}
\widetilde{\omega}^{0}_{5}(\mathcal{A},\mathcal{F}_{V},
\mathcal{F}_{A})&=2{\rm Tr\,}\Big(3\widehat{\mathcal{A}}\widehat{\mathcal{F}}_{V}-2\widehat{\mathcal{A}}^{\,3}\Big)({\rm Tr\,}\mathcal{F}_{V}).
\end{align}
This expression makes it explicit that two terms come respectively from the SU(2)$^{2}$U(1) triangle and the SU(2)$^{3}$U(1) square diagrams.
Playing the same game with the traces in Eq. \eqref{eq:omega14AV_basis}, we retrieve the expression leading to the anomaly found 
in Ref.~\cite{Kaiser:2000ck} for two flavor QCD
\begin{align}
\widetilde{\omega}^{1}_{4}(u_{A},\mathcal{A},\mathcal{F}_{V})&=6{\rm Tr\,}\Big[\widehat{u}_{A}\Big(\widehat{\mathcal{F}}_{V}
-2\widehat{\mathcal{A}}^{\,2}\Big)\Big]({\rm Tr\,}\mathcal{F}_{V}),
\end{align}
where, by consistency, the axial-vector gauge function $u_{A}$ is taken not to have components on U(1)$_{A}$.


\section{Chern-Simons effective actions, currents, and anomaly inflow}
\label{sec:CS}

In Section \ref{sec:LRsymm_anom} we have seen how the anomalous effective action $\Gamma[\mcA]_{\rm CS}$
can be constructed from the Chern-Simons form $\omega^{0}_{2n-1}$ [see Eq. \eqref{eq:CS_eff_action}]. In the next section, we 
show how this effective action can be used to generate an anomalous partition function 
capturing the physical consequences of the anomaly under stationary conditions. But before that, in order to understand how this is possible, 
we need to study the currents induced by $\Gamma[\mcA]_{\rm CS}$. These are obtained by applying 
a general variation of the gauge field that will be denoted $\delta\mcA=B$, with $B$ an infinitesimal Lie-algebra valued one-form.

The result of this variation on the Chern-Simons form, $\delta_B \omega^{0}_{2n-1}(\mcA)$, 
can be efficiently computed with the help of the generalized transgression formula (see \cite{Manes:1985df} and Appendix \ref{app:transgression}). More specifically 
\begin{align}
\omega^{0}_{2n-1}(\mcA+B)-\omega^{0}_{2n-1}(\mcA)=\int_0^1 \ell_t\,d\omega^{0}_{2n-1}(\mcA_t)+d\int_0^1 \ell_t\,\omega^{0}_{2n-1}(\mcA_t)
\label{eq:genhom1}
\end{align}
where $\mcA_t=\mcA + tB$ is a family of connections interpolating between $\mcA$ and $\mcA+B$, while the action of the operator $\ell_t$ is given by 
\begin{align}
\ell_{t}\mcA_{t}&=0, \nonumber \\[0.2cm]
\ell_t \mcF_t&=d_t\mcA_t=dtB.
\end{align} 
Notice that, in order to compute the currents we only need to evaluate \eqref{eq:genhom1} to linear order in $B$, using 
that for any function of $\mcA$ and $\mcF$ 
\begin{align}
\int_0^1 \ell_tf(\mcA_t,\mcF_t)=\ell f(\mcA,\mcF)+\mathcal{O}(B^2), 
\end{align}
where the operator $\ell$, introduced in~\cite{Bardeen:1984pm}, is defined by $\ell\mcF=B$, $\ell\mcA=0$.

Thus, to linear order in $B$,  the  first term in the right-hand side of \eqref{eq:genhom1} evaluates to
\begin{align}
\int_0^1 \ell_t d\omega^{0}_{2n-1}(\mcA_t)&=\int_0^1 \ell_t{\rm Tr\,}\mcF_t^{n} 
=\ell\,{\rm Tr\,}\mcF^{n}+\mathcal{O}(B^{2})\nonumber \\[0.2cm]
&=n{\rm Tr\,}(B\mcF^{n-1})+\mathcal{O}(B^{2}),
\label{eq:first_term}
\end{align}
while the second one gives
\begin{align}
\int_0^1 \ell_t\,\omega^{0}_{2n-1}(\mcA_t)= \ell\,\omega^{0}_{2n-1}(\mcA)+\mathcal{O}(B^{2}).
\end{align}
Integrating Eq. (\ref{eq:genhom1}) over $\mcM_{2n-1}$ and using the Stokes theorem finally yields the following expression for the variation of the effective 
action functional
\begin{align}
\delta_B\Gamma[\mcA]_{\rm CS}&=\int\limits_{\mathcal{M}_{2n-1}}\ell\,\mathcal{P}_{n}(\mcF)
+c_{n}\int\limits_{\mcM_{2n-2}}\ell\omega^{0}_{2n-1}(\mcA) \nonumber \\[0.2cm]
&\equiv\int\limits_{\mcM_{2n-1}} \, {\rm Tr\,}(B\mcJ_{\rm bulk})-\int\limits_{\mcM_{2n-2}} 
{\rm Tr\,}(B\mcJ_{\rm BZ}),
\label{eq:CS_currents}
\end{align}
The bulk current term on the second line can be read from \eqref{eq:first_term}
\begin{align}
{\rm Tr\,}(B\mcJ_{\rm bulk})=\ell\,\mathcal{P}_{n}(\mcF) \hspace*{0.5cm} \Longrightarrow \hspace*{0.5cm}
\mcJ_{\rm bulk}=nc_{n}\mcF^{n-1}.
\label{eq:JBulk}
\end{align}
On the other hand, the action of the operator $\ell$ on the Chern-Simons form gives the covariant (Bardeen-Zumino) 
current~\cite{Bardeen:1984pm}
\begin{align}
c_n \ell\,\omega^{0}_{2n-1}(\mcA)=-{\rm Tr\,}(B\mcJ_{\rm BZ}).
\label{eq:BZhom}
\end{align}
In fact, $\mcJ_{\rm BZ}$ is a $(2n-3)$-form taking values on the Lie algebra $\mathfrak{g}$, dual to the Bardeen-Zumino one-form 
current which added to the consistent current gives the covariant one~\cite{Bardeen:1984pm}. Henceforth, all currents will be represented by the corresponding dual forms (see the Appendix \ref{app:formulae} for details).

The restriction of $\mcJ_{\rm bulk}$ to the boundary $\mcM_{2n-1}$ turns out to coincide with  (minus) the covariant anomaly, in agreement with the anomaly inflow mechanism~\cite{Callan:1984sa,Jensen:2013kka,Haehl:2013hoa}. To see this, we go back to Eq. \eqref{eq:jcons} and
evaluate \eqref{eq:CS_currents} for a gauge variation of the gauge field, i.e., we set
\begin{align}
B\equiv \delta_u\mcA=du+[\mcA,u]\equiv Du\,.
\end{align}
Integrating then by parts and using the cyclic property of the trace, together with the Stokes theorem applied on 
the first term in the right-hand side of \eqref{eq:CS_currents}, we have
\begin{align}
\int\limits_{\mcM_{2n-1}} \, {\rm Tr\,}\Big[(Du)\mcJ_{\rm bulk}\Big]
=-\int\limits_{\mcM_{2n-1}} \, {\rm Tr\,}\Big[u(D\mcJ_{\rm bulk})\Big]
+\int\limits_{\mcM_{2n-2}} \, {\rm Tr\,}\Big(u\mcJ_{\rm bulk}\Big),
\end{align}
while the second term can be written as
\begin{align}
-\int\limits_{\mcM_{2n-2}} \, {\rm Tr\,}\Big[(Du)\mcJ_{\rm BZ}\Big]
=\int\limits_{\mcM_{2n-2}} \, {\rm Tr\,}\Big[u(D\mcJ_{\rm BZ})\Big].
\end{align}
Finally, noting that the Bianchi identity $D\mcF=0$ implies $D\mcJ_{\rm bulk}=0$, we find that the gauge variation of the Chern-Simons action can be written
\begin{align}
\delta_u\Gamma[\mcA]_{\rm CS}&=\int\limits_{\mcM_{2n-2}} \, {\rm Tr\,}\Big[u(\mcJ_{\rm bulk}+D\mcJ_{\rm BZ})\Big] 
\nonumber \\[0.2cm]
&=-\int\limits_ {\mcM_{2n-2} }{\rm Tr\,}\Big(u\,G[\mcA]_{\rm cons}\Big),
\end{align}
where we have used \eqref{eq:noninv} to get the last equality. This implies
\begin{align}
\Bigl.\mcJ_{\rm bulk}\Bigr|_{\mcM_{2n-2} }=-\Big(G[\mcA]_{\rm cons}+D\mcJ_{\rm BZ}\Big)=-G[\mcA]_{\rm cov},
\end{align}
where $G[\mcA]_{\rm cov}$ is the covariant anomaly.
This completes our identification of the currents induced by the Chern-Simons effective action.

Using Eq. \eqref{eq:BZhom} for $n=3$, we retrieve the Bardeen-Zumino current for a right-handed fermion in four dimensions
\begin{align}
\mcJ_{\rm BZ}^R={i\over 24\pi^{2}}\left(\mcF_R\mcA_R+\mcA_R\mcF_R-\frac{1}{2}\mcA_R^3\right),
\label{eq:BZ4}
\end{align}
while the left-handed current differs just by a relative minus sign. 
It will be useful to write these expressions also in terms of vector and axial gauge fields. 
Thus, instead of \eqref{eq:BZhom} we use 
\begin{align}
c_n \ell\,\widetilde\omega_{2n-1}^{0}(\mcV,\mcA)=-{\rm Tr\,}\Big(B_V\mcJ^V_{\rm BZ}+B_A\mcJ^A_{\rm BZ}\Big),
\label{eq:BZVA}
\end{align}
where $B_{V}$ and $B_{A}$ are the shifts of the vector and axial-vector gauge fields respectively. The operator 
$\ell$ annihilate the gauge fields, $\ell\mcV=\ell\mcA=0$, whereas its action on the field strengths is defined 
by $\ell\mcF_V=B_V$ and $\ell\mcF_A=B_A$. Thus, the Bardeen-Zumino currents in four dimensions are given by
\begin{align}
\mathcal{J}^V_{\rm BZ} &= {i\over 12\pi^{2}} \Big[3(\mcF_V \mcA+\mcA\mcF_V)-4\mcA^3\Big],    \\[0.2cm]
 \mathcal{J}^A_{\rm BZ} & = {i\over 12\pi^{2}}\Big(\mcF_A \mcA+\mcA\mcF_A  \Big).
\end{align} 
For the bulk currents, we generalize \eqref{eq:JBulk} to
\begin{align}
\ell\,\mathcal{P}_{n}(\mcV,\mcA)={\rm Tr\,}\Big(B_V\mcJ^V_{\rm bulk}+B_A\mcJ^A_{\rm bulk}\Big),
\end{align}
which for $n=3$ gives the Bardeen form of the bulk currents
\begin{equation}
\begin{split}
\mathcal{J}^V_{\rm bulk} & = -{i\over 24\pi^{2}} \Big( \mcF_V ^2+\mcF_A ^2  \Big),    \\[0.2cm]
 \mathcal{J}^A_{\rm bulk} & = -{i\over 24\pi^{2}}\Big(\mcF_V\mcF_A +  \mcF_A \mcF_V  \Big)  \, .
 \label{eq:Jbulk}
\end{split}
\end{equation} 

At this point it is important to stress that the difference between the left-right symmetric 
and Bardeen forms of the consistent anomaly stems from the use of different Chern-Simons forms, i.e., (\ref{eq:CS_Sym}) and (\ref{om5_vector}) respectively. 
On the other hand, the invariant polynomial is unique and so are the bulk currents that, restricted to physical spacetime, give the covariant anomalies. In other words, the covariant anomalies in both the left-right symmetric 
and Bardeen forms are exactly the same. The consequence is that there is no way of constructing a conserved covariant vector current.

\section{The anomalous partition function from dimensional reduction}
\label{sec:anomalousPF}

We turn now to the computation of
the partition function of anomalous hydrodynamics using the differential geometry formalism for quantum anomalies 
reviewed in Sec.~\ref{sec:anom}. 
Given the anomalous effective functional, the partition function $\log Z$ can be obtained by using the
imaginary time prescription:
we take all fields to be time-independent and compactify the Euclidean time direction to a circle of 
length $\beta$. This amounts to replacing the integration over time in 
\hbox{$\log Z \equiv W[\mcA]= i \Gamma[ \mcA]$} according to
\begin{align} 
i\int dx^0 \longrightarrow  \beta_0\equiv {1\over T_{0}}, 
\label{eq:it_prescription}
\end{align}
with $T_0$ the equilibrium temperature. 

On the other hand, we  know that the gauge variation of the Chern-Simons action $\Gamma[\mcA]_{\rm CS}$ defined in Eq. 
\eqref{eq:CS_eff_action}
reproduces the consistent anomaly. Thus we may expect the Chern-Simons action evaluated on a time-independent background to be closely related to the anomalous partition function for that background. One potential difficulty is that $\Gamma[\mcA]_{\rm \rm CS}$ lives in a five-dimensional manifold $\mcM_{5}$ and we saw in the previous section that the boundary current induced on four-dimensional spacetime $\mcM_{4}$\,=\,$\partial\mcM_{5}$ is the Bardeen-Zumino current, \emph{not}  the consistent current  one would expect from an effective functional [cf. Eq. (\ref{eq:jcons})]. Nevertheless, we will see that the dimensional reduction of the Chern-Simons action on a time-independent background yields a satisfactory anomalous partition function.

As explained in the Introduction, we want to consider a gauge theory on a stationary background. Choosing an 
appropriate gauge, its line element can be written as~\cite{Banerjee:2012iz}
\begin{align}
ds^2 &= - e^{2\sigma(\x)} \Big[dt + a_{i}(\x) dx^{i}\Big]^2 + g_{ij}(\x)dx^i dx^j,
\label{eq:metric}
\end{align}
where all ten metric functions $\{\sigma(\x),a_{i}(\x),g_{ij}(\x)\}$ are independent of the time coordinate~$t$.
This metric remains invariant under time-independent shifts of the time coordinate, combined with the appropriate transformation of
the metric functions $a_{i}(\x)$. Using the notation of Refs.~\cite{Banerjee:2012iz,Bhattacharyya:2012xi}
\begin{align}
t &\longrightarrow t'=t+\phi(\x), \nonumber \\[0.2cm]
\x &\longrightarrow \x'=\x, \label{eq:KK_trans} \\[0.2cm]
a_{i}(\x) &\longrightarrow a'_{i}(\x)= a_{i}(\x)-\partial_{i}\phi(\x). 
\nonumber
\end{align}
For obvious reasons, this isometry is referred to as Kaluza-Klein (KK) gauge transformations.
In a theory with static gauge fields, their components
\begin{align}
\mcA_{\mu}=\big(\mcA_{0}(\mathbf{x}),\mcA_{i}(\mathbf{x})\big),
\end{align}
transform under \eqref{eq:KK_trans} according to
\begin{align}
\mcA_{0}(\x) &\longrightarrow \mcA_{0}(\x), \nonumber \\[0.2cm]
\mcA_{i}(\x)&\longrightarrow \mcA_{i}'(\x)=\mcA_{i}(\x)-\mcA_{0}(\x)\partial_{i}\phi(\x).
\end{align}
This implies the invariance of the following combinations
\begin{align}
A_{i}(\x)&\equiv \mcA_{i}(\x)-\mcA_{0}(\x)a_{i}(\x),
\label{eq:boldA_form_def}
\end{align}
so $A_{\mu}=(\mcA_{0},A_{i})$ denote the KK-invariant gauge fields, which on the other hand behave as standard gauge fields
under time-independent gauge transformations. The use of the gauge \eqref{eq:metric}, together with the systematic implementation of 
KK invariance, leads to explicit results that can be easily applied to hydrodynamics, as shown in Ref.~\cite{Banerjee:2012iz}.

In the language of differential forms, the original gauge field
can be decomposed into KK-invariant quantities by writing 
\begin{align}
\mcA(\x)&=\mcA_0(\x) \Big[dx^0+a_i(\x)dx^i\Big]+\Big[\mcA_i(\x)-\mcA_0(\x)a_i(\x)\Big]dx^i \nonumber \\[0.2cm]
&=\mcA_0(\x)\theta(\x) +A_i(\x)dx^i \equiv  \mcA_0(\x)\theta(\x) +   \bm{A}(\x),
\label{eq:KKdecomp}
\end{align}
where we have defined the  one-forms 
\begin{align}
a(\x)&=a_i(\x)dx^{i}, \nonumber \\[0.2cm]
\theta(\x)&=dx^0+a(\x). 
\label{eq:def_a_theta}
\end{align}
Note that the one-forms $\mcA_0$, $\theta$, 
and $\mbA$ all remain invariant under KK transformations. The corresponding decomposition of the field strength in terms of these 
quantities reads
\begin{align}
\mcF=\mbD(\mcA_{0}\theta)+\mbF,
\label{eq:field_strenght_decomp_theta_da}
\end{align}
where $\mbD$ is the covariant derivative associated with $\mbA$, acting on $p$-forms according to 
\begin{align}
\mbD\omega_{p}\equiv d\omega_{p}+\mbA\omega_{p}-(-1)^{p}\omega_{p}\mbA,
\label{eq:covariant_derivative_bfA}
\end{align}
and we have introduced the field strength associated to $\mbA$
\begin{align}
\mbF=d\mbA+\mbA^{2}.
\end{align}

Incidentally, these metric and field theoretical functions are related to hydrodynamic quantities such as 
the local temperature $T(\x)$, the chemical potential
$\mu(\x)$, and the fluid 
velocity $u^{\mu}(\x)$ by~\cite{Banerjee:2012iz,Jensen:2013kka}
\begin{align}
T(\x)&=T_{0}\,e^{-\sigma(\x)}, \nonumber \\[0.2cm]
\mu(\x)&=e^{-\sigma(\x)}\mcA_{0}(\x), \label{eq:fluid_variables1}\\[0.2cm]
u(\x)&\equiv u_{\mu}(\x)dx^{\mu}= -e^{\sigma(\x)}\theta(\x). \nonumber
\end{align}
In terms of them, the decomposition \eqref{eq:KKdecomp} reads
\begin{align}
\mcA&=\mbA-\mu u.
\label{eq:A=A-muu}
\end{align}
Thus, our field $\mbA$ corresponds to the hatted connection of Ref. \cite{Jensen:2013kka}. Moreover, from the identity
$du(\x)=-u(\x)\mathfrak{a}(\x)+2\omega(\x)$, we identify the acceleration $\mathfrak{a}(\x)$ and vorticity $\omega(\x)$ to be
\begin{align}
\mathfrak{a}(\x)&=d\sigma(\x), \nonumber \\[0.2cm]
\omega(\x)&=-{1\over 2}e^{\sigma(\x)}da(\x).
\label{eq:fluid_variables2}
\end{align}

After all these prolegomena, we are in position to compute the anomaly-induced equilibrium partition function implementing 
dimensional reduction. Taking into account the decomposition \eqref{eq:KKdecomp}, we can write the identity
\begin{align}
\int\limits_{\mcM_{2n-1} } \omega^{0}_{2n-1}(\mcA)=\int\limits_{\mcM_{2n-1} } \omega^{0}_{2n-1}(\mcA_0\theta+\mbA)-\int\limits_{\mcM_{2n-1} }  \omega^{0}_{2n-1}( \mbA),
\end{align}
which trivially holds since $\mbA$ is independent of $dx^0$ and the last integral vanishes upon dimensional reduction. The advantage of 
writing the effective action in this fashion is that 
the combination of terms on the right-hand side of this equation
can be evaluated with the help of the same generalized homotopy formula already used in \eqref{eq:genhom1}, namely
\begin{align}
\omega^{0}_{2n-1}(\mcA_0\theta +   \mbA)-\omega^{0}_{2n-1}(\mbA)=\int_0^1 \ell_t\,d\omega^{0}_{2n-1}(\mcA_t)+d\int_0^1 \ell_t\,\omega^{0}_{2n-1}(\mcA_t)\,,
\label{eq:genhom2}
\end{align}
where now we consider the family of connections $\mcA_t=\mbA+t \mcA_0\theta$. The details of the computation are given in Appendix \ref{app:transgression}, where it is shown that the first term evaluates to
\begin{align}
\int_0^1 \ell_t\,d\omega^{0}_{2n-1}(\mcA_t)= n\int_0^1 dt \,dx^0\, {\rm Tr\,} \Big[\mcA_0(\mbF+t\mcA_0da)^{n-1}\Big],
\label{eq:integral_domega0_2n-1stationary}
\end{align}
while the second one yields
\begin{align}
\int_0^1 \ell_t\,\omega^{0}_{2n-1}(\mcA_t)=\int_0^1 dt\, dx^0
\left. \mcA_0\frac{\delta}{\delta\mcF}\omega^{0}_{2n-1}(\mcA,\mcF)\right|_{\substack{\hspace*{-1cm}\mcA\to\mbA \\[0.1cm]\mcF\to\mbF+t\mcA_0da}}.
\label{eq:integral_domega0_2n-1stationary_2ndterm}
\end{align}

Substituting  these results into Eq. \eqref{eq:CS_eff_action} we arrive at a very interesting decomposition of the 
Chern-Simons effective action $\Gamma[\mcA_0,\mbA]$. Indeed, after implementing the prescription \eqref{eq:it_prescription},
we find that the partition function naturally splits into two terms \cite{Jensen:2013kka}
\begin{align}
i\Gamma[\mcA_0,\mbA]_{\rm CS}=W[\mcA_0,\mbA]_{\rm inv}+W[\mcA_0,\mbA]_{\rm anom}.
\label{eq:dired}
\end{align}
We have found that the two terms are explicitly given by
\begin{align}
W[\mcA_0,\mbA]_{\rm inv}=\frac{nc_n}{T_0}\int\limits_{D_{2n-2}}\int_0^1 dt\,{\rm Tr\,}\Big[\mcA_0(\mbF+t\mcA_0da)^{n-1}\Big]
\label{eq:Winv}
\end{align}
and
\begin{align}
W[\mcA_0,\mbA]_{\rm anom}=\left.\frac{c_n}{T_0}\int\limits_{S^{2n-3}} \int_0^1 dt \mcA_0\frac{\delta}{\delta\mcF}\omega^{0}_{2n-1}(\mcA,\mcF)\right|_{\substack{\hspace*{-1cm}\mcA\to\mbA \\[0.1cm]\mcF\to\mbF+t\mcA_0da}}.
\label{eq:Wanom}
\end{align}
Note that the imaginary time formalism requires to assume that $\mcM_{2n-1}=D_{2n-2}\times S^1$ and $\mcM_{2n-2}=S^{2n-3}\times S^1$, where $\mcM_{2n-2}=\partial D_{2n-1}$ and $S^1$ denotes the thermal cycle of length $\beta=T_{0}^{-1}$. Inspection of Eqs. \eqref{eq:Winv} and 
\eqref{eq:Wanom} shows that whereas $W_{\rm inv}$ is of order $n-1$ in the derivative expansion, the order of the local anomalous piece 
$W_{\rm anom}$ is $n-2$.

A first important thing to be noticed is that $W[\mcA_0,\mbA]_{\rm inv}$ is given as an integral over a $(2n-1)$-dimensional Euclidean manifold, and is therefore  nonlocal from the viewpoint of $(2n-2)$-dimensional spacetime. However, $W[\mcA_0,\mbA]_{\rm inv}$ is manifestly invariant under time-independent gauge transformations and therefore does not contribute to the gauge anomaly. As a consequence, it is the second piece $W[\mcA_0,\mbA]_{\rm anom}$ which completely accounts for the gauge noninvariance of the partition function $i\Gamma[\mcA_0,\mbA]_{\rm CS}$ and its gauge variation reproduces the consistent anomaly. As $W[\mcA_0,\mbA]_{\rm anom}$ is given as  a local integral over the spatial manifold $S^{2n-3}$,  the rationale behind Eqs. (\ref{eq:jcons})-(\ref{eq:consanom})  shows  that it must induce the consistent current for the time-independent background. This makes $W[\mcA_0,\mbA]_{\rm anom}$ wholly satisfactory as an anomalous partition function.  Notice as well that, although the integral of the Chern-Simons form $\omega^{0}_{2n-1}$ is a 
topological invariant and therefore does not depend on the background metric, after dimensional reduction both the 
invariant and anomalous part of the effective action pick up a dependence on the metric function $a_{i}$. From 
Eqs. \eqref{eq:Winv} and \eqref{eq:Wanom}, we see that this dependence always comes through the KK-invariant combination $da$.

In four-dimensional spacetime ($n=3$), the anomalous, local piece of the partition function $W[\mcA_0,\mbA]_{\rm anom}$ is 
explicitly given by
\begin{align}
W[\mcA_0,\mbA]_{\rm anom}=-{i\over 24\pi^{2} T_0}\int\limits_{S^{3}} {\rm Tr\,}\left[   \mcA_0\left(\mbF\mbA+\mbA\mbF-\frac{1}{2}\mbA^3\right) +\mcA_0^2\,\mbA da   \right]\,.
\label{eq:Wanom4}
\end{align}
It is very important to keep in mind that the fact that $W[\mcA_0,\mbA]_{\rm inv}$ is not contributing to the anomaly does not mean that it can be discarded as a useless byproduct of dimensional reduction. As we will see in the next section, the most efficient way to compute the covariant currents and the energy-momentum tensor involves $W[\mcA_0,\mbA]_{\rm inv}$. In a four-dimensional spacetime this term reads
\begin{align}
W[\mcA_0,\mbA]_{\rm inv}=-{i\over 8\pi^{2} T_0}\int\limits_{D_4} {\rm Tr\,}\left[\mcA_0\mbF^2+da\,\mcA_0^2\,\mbF +\frac{1}{3}(da)^2\mcA_0^3\right]\,.
\label{eq:Winv4D}
\end{align}
It should be stressed again that $D_4$ is a four-dimensional Euclidean hypersurface containing an extra spatial coordinate, so it is nonlocal from the viewpoint of four-dimensional spacetime.

With applications to hadronic fluids in mind, it is convenient to write the Bardeen form of the anomalous part of the partition function in terms of vector and axial gauge fields $\mcV_{\mu}$ and $\mcA_{\mu}$, whose components
are denoted respectively by
\begin{align}
\mcV_\mu &= \big(\mcV_0(\x),\mcV_i(\x)\big), \nonumber \\[0.2cm] 
\mcA_\mu &= \big(\mcA_0(\x),\mcA_i(\x)\big). 
\end{align}
Again, we can define the new gauge fields
\begin{align}
V_{\mu}(\x)&\equiv \big(V_{0}(\x), V_{i}(\x)\big)=\big(\mcV_{0}(\x),\mcV_{i}(\x)-V_{0}(\x)a_{i}(\x)\big), \nonumber \\[0.2cm]
A_{\mu}(\x)&\equiv \big(A_{0}(\x), A_{i}(\x)\big)=\big(\mcA_{0}(\x),\mcA_{i}(\x)-A_{0}(\x)a_{i}(\x)\big).
\end{align}
which remain invariant under KK transformations \eqref{eq:KK_trans}. 
Repeating the analysis leading to Eq. \eqref{eq:Wanom}, we find in this case
\begin{align}
W[\mcV_0,\mcA_0,\mbV,\mbA]_{\rm anom}&=
\frac{c_n}{T_0}\int_{S^{2n-3}}\int_0^1 dt
\left\{\,
\left[\left(\mcV_{0}{\delta\over \delta\mcF_V}\right)
\widetilde{\omega}^{0}_{2n-1}(\mcV,\mcA)\right]
\Bigg|_{\blacksquare\rightarrow {\blacksquare}_{t}}\right. \nonumber \\[0.4cm]
&+\left.\left[\left(\mcA_{0}{\delta\over \delta\mcF_A}\right)
\widetilde{\omega}^{0}_{2n-1}(\mcV,\mcA)\right]
\Bigg|_{\blacksquare\rightarrow{\blacksquare}_{t}}\right\},
\label{eq:WanomVA}
\end{align}
where we have used the compact notation
\begin{align}
\blacksquare\rightarrow {\blacksquare}_{t}:\left\{
\begin{array}{l}
\mcV\longrightarrow \mbV, \\[0.2cm] \mcA\longrightarrow \mbA\\[0.2cm]
\mcF_V\longrightarrow\mbF_V+tV_{0}da, \\[0.2cm] \mcF_A\longrightarrow \mbF_A+tA_{0}da\\
\end{array}
\right.\hspace*{0.4cm}.
\end{align}
Evaluating this expression for $n=3$, we find the anomalous partition function in four dimensions to be
\begin{align}
W[\mcV_0,\mcA_0,\mbV,\mbA]_{\rm anom}
&=-{i\over 4\pi^{2} T_0}\int\limits_{S^{3}} {\rm Tr\,}\Bigg[   
\mcV_0 \left(\mbF_V \mbA+\mbA\mbF_V-{4\over 3}\mbA^3\right)
\nonumber \\[0.2cm]
&+{1\over 3}\mcA_0\Big(\mbF_A\mbA+\mbA\mbF_A\Big)+{1\over 3}da\Big(3\mcV_0^2+\mcA_0^2\Big)\mbA
 \Bigg].
\label{eq:WanomVA4}
\end{align}

On the other hand, $W[\mcV_0,\mcA_0,\mbV,\mbA]_{\rm inv}$ is directly constructed  from the corresponding 
invariant polynomial and is therefore the same for both the left-right symmetric and  Bardeen form of the anomaly. 
This means that $W[\mcV_0,\mcA_0,\mbV,\mbA]_{\rm inv}$ can be simply obtained by performing a mere change of variables on $W[\mcA_{R0},\mbA_R]_{\rm inv}-W[\mcA_{L0},\mbA_L]_{\rm inv}$. The resulting partition function in four dimensions 
($n=3$) is
\begin{align}
W[\mcV_0,\mcA_0,\mbV,\mbA]_{\rm inv}=&\;\;-{i\over 4\pi^{2} T_0}\int\limits_{D_4} {\rm Tr\,}\Bigg\{  
\mcA_0\Big(\mbF_V^2+\mbF_A^2\Big)+\mcV_0\Big(\mbF_V\mbF_A+\mbF_A\mbF_V\Big)  \nonumber\\
&+da\Big[\Big(\mcV_0^2+\mcA_0^2\Big)\mbF_A+\Big(\mcV_0\mcA_0+\mcA_0\mcV_0\Big)\mbF_V\Big] \nonumber\\[0.2cm] 
&+\frac{1}{3}(da)^2\Big(\mcA_0^3+3\mcA_0\mcV_0^2\Big)
\Bigg\}.
\label{eq:WinvVA4}
\end{align}
Equations~\eqref{eq:WanomVA4} and \eqref{eq:WinvVA4} are the main results of this section.

We can particularize the anomalous partition function \eqref{eq:WanomVA4} for the gauge group
U(2)$_{L}\times$U(2)$_{R}$, relevant for a two-flavor hadronic fluid. 
In this case, the four generators $t_{a}$ (with $a=0,1,2,3$) are given by
\begin{align}
t_{0}={1\over 2}\mathbb{1}, \hspace*{1cm} t_{i}={1\over 2}\sigma_{i}, 
\label{eq:generatorsU(2)}
\end{align}
with $\sigma_{i}$ the Pauli matrices ($i=1,2,3$). Due to the properties of the U(2) generators, most 
traces factorize  (see Appendix \ref{app:traces}) and
the anomalous part of the effective
action takes the form
\begin{align}
W[\mcV_0,\mcA_0,\mbV,\mbA]_{\rm anom}^{{\rm U}(2)\times{\rm U}(2)}
&=-{i\over 4\pi^{2} T_0}\int\limits_{S^{3}} \left\{   
\big({\rm Tr\,}\mcV_0\big){\rm Tr\,}\left(\mbF_V \mbA-{2\over 3}\mbA^3\right)
+\big({\rm Tr\,}\mbF_{V}\big){\rm Tr\,}\big(\mcV_{0}\mbA\big)
\right.\nonumber \\[0.2cm]
&+\Big[{\rm Tr\,}\big(\mcV_{0}\mbF_{V}\big) 
-({\rm Tr\,}\mcV_{0})({\rm Tr\,}\mbF_{V})\Big]({\rm Tr\,}\mbA)+{1\over 3}({\rm Tr\,}\mbF_{A}){\rm Tr\,}(\mcA_{0}\mbA)
\nonumber \\[0.2cm]
&+{1\over 3}\Big[{\rm Tr\,}(\mcA_{0}\mbF_{A})
-({\rm Tr\,}\mcA_{0})({\rm Tr\,}\mbF_{V})\Big]({\rm Tr\,}\mbA)+{1\over 3}({\rm Tr\,}\mcA_{0}){\rm Tr\,}(\mbF_{A}\mbA)  \nonumber \\[0.2cm]
&-{2\over 3}\big({\rm Tr\,}\mbA\big){\rm Tr\,}\big(\mcV_{0}\mbA^{2}\big)-{1\over 2}
da\Big[({\rm Tr\,}\mcV_{0})^{2}-{\rm Tr\,}\mcV_{0}^{2}\Big]({\rm Tr\,}\mbA)
\label{eq:W_anom_VA_U2xU2} \\[0.2cm]
&-{1\over 6}da\Big[({\rm Tr\,}\mcA_{0})^{2}-{\rm Tr\,}\mcA_{0}^{2}\Big]({\rm Tr\,}\mbA)+da({\rm Tr\,}\mcV_{0}){\rm Tr\,}(\mcV_{0}\mbA)
\nonumber \\[0.2cm]
&+{1\over 3}da({\rm Tr\,}\mcA_{0}){\rm Tr\,}(\mcA_{0}\mbA) \bigg\}.
\nonumber
\end{align} 
The expression gets much simpler when axial-vectors fields do not have components on the U(1)$_{A}$ factor, so their traces vanish. In this
case, we have
\begin{align}
W[\mcV_0,\mcA_0,\mbV,\mbA]_{\rm anom}^{{\rm U}(2)\times{\rm U}(2)}
&=-{i\over 4\pi^{2} T_0}\int\limits_{S^{3}} \left\{   
\big({\rm Tr\,}\mcV_0\big){\rm Tr\,}\left(\mbF_V \mbA-{2\over 3}\mbA^3\right)
+\big({\rm Tr\,}\mbF_{V}\big){\rm Tr\,}\big(\mcV_{0}\mbA\big)
\right.\nonumber \\[0.2cm]
&+da\big({\rm Tr\,}\mcV_{0}\big){\rm Tr\,}(\mcV_{0}\mbA)\bigg\}.
\label{eq:Wanomnoaxialtrace}
\end{align}
This form of the partition function accounts for the two-flavor anomalies of QCD when 
the global  chiral symmetry SU(2)$_L \times $SU(2)$_R \times$ U(1)$_V$ is not broken, 
but as we will show, it may also be used to obtain the anomalous partition function  
when a SU(2) multiplet of  pions  is included.
An inspection of this expression reveals its diagrammatic 
origin: the terms emerge from  
the U(1)SU(2)$^{2}$ triangle and the U(1)SU(2)$^{3}$ square diagrams. 
The latter are required to guarantee the Wess-Zumino consistency equations,  
although previous studies~\cite{Neiman:2010zi} only considered the effects of anomalous triangle diagrams 
through $\text{Tr}\bigl(t_a \{t_b, t_c\}\bigr)$.


\section{Gauge currents and  energy-momentum tensor in stationary backgrounds}
\label{sec:currents_em_tensor}

Once we have arrived at a general prescription to obtain the equilibrium partition function, 
we undertake the construction of the anomaly-induced consistent and covariant gauge currents, as well as the energy-momentum tensor. 
By expressing them in terms of the appropriate 
fluid fields listed in 
Eqs.~\eqref{eq:fluid_variables1} and~\eqref{eq:fluid_variables2}, the hydrodynamic constitutive relations can be obtained. 

\subsection{Consistent and covariant currents}

From Eq. \eqref{eq:CS_currents}, we know that the physical spacetime current induced by the Chern-Simons 
action $\Gamma[\mcA]_{\rm CS}$  equals  (minus) the Bardeen-Zumino current.
On the other hand, we have argued that $W[\mcA_0,\mbA]_{\rm anom}$, which is  the boundary component of the dimensional reduction of 
$i\Gamma[\mcA]_{\rm CS}$ on a time-independent background, induces the consistent current. These two results can be mutually consistent only if the boundary current $\mathcal{X}$ induced by $W[\mcA_0,\mbA]_{\rm inv}$
\begin{align}
\delta_{B}W[\mcA_0,\mbA]_{\rm inv}=\int\limits_{S^{2n-3}}{\rm Tr\,}\Big(B\mathcal{X}\Big)+\mbox{bulk contribution},
\end{align}
satisfies
\begin{align}
\mcJ_{\rm cons}+\mathcal{X}=-\mcJ_{\rm BZ} \hspace*{0.5cm}
\Longrightarrow \hspace*{0.5cm} \mathcal{X}=-(\mcJ_{\rm cons}+\mcJ_{\rm BZ})=-\mcJ_{\rm cov}.
\end{align}
This shows that  the physical spacetime current induced by $W[\mcA_0,\mbA]_{\rm inv}$   is (minus) the covariant current. 
Exploiting this fact, we can obtain a very simple expression for $\mcJ_{\rm cov}$, as shown in the following.

Let us analyze this issue in more detail. 
From \eqref{eq:KKdecomp}, a general variation of the gauge field $\mcA$ admits the following Kaluza-Klein invariant decomposition
\begin{align}
B\equiv \delta_{B}\mcA=\delta_{B}\mcA_0\theta+\delta_{B}\mbA\equiv \mcB_0\theta+\mbB
\label{eq:B_general_varA}
\end{align}
where $\mcB_0$ and $\mbB$ are respectively conjugate to the KK invariant currents \hbox{$\mcJ_0$ and $\mbJ$} (see~\cite{Banerjee:2012iz} and Appendix \ref{app:formulae} for details). 
As shown in \cite{Jensen:2013kka}, the covariant version of these currents, \hbox{$\mcJ_{0,\rm cov}$ and $\mbJ_{\rm cov}$}, can be obtained 
from the variation of the nonanomalous part of the effective action. Given that this is a nonlocal functional in $2n-2$ dimensions,
it is enough to extract the terms proportional to $d\mcB_0$ and $d\mbB$ from a general variation \eqref{eq:B_general_varA}, since these are the only terms that will give boundary contributions upon integration by parts of  $\delta_{B}W[\mcA_0,\mbA]_{\rm inv}$. Now, from the general expression for this functional given in Eq. \eqref{eq:Winv}, it is obvious that $\delta_B W[\mcA_0,\mbA]_{\rm inv}$ is independent of $d\mcA_0$,  with the  consequence  that $\mcJ_{0,\rm cov}=0$. This fact notwithstanding, we will see in the next section that this is no longer true in the presence of spontaneous symmetry breaking, where we will find a nonvanishing anomalous contribution to $\mcJ_{0,\rm cov}$.

On the other hand, the only dependence of $\delta_{B}W[\mcA_0,\mbA]_{\rm inv}$ on $d\mbB$ is through 
\begin{align}
\delta_{B}\mbF=d\mbB+\{\mbA,\mbB\}.
\end{align}
Thus, integrating by parts $\delta_{B}W[\mcA_0,\mbA]_{\rm inv}$, and taking into account a minus sign from the 
Stokes theorem, yields the following
 expression for the anomalous covariant current in a stationary background (cf. \cite{Jensen:2013kka})
\begin{align}
\mcJ_{0,\rm cov}&=0, \nonumber \\[0.2cm]
\mbJ_{\rm cov}&=T_0\frac{\delta }{\delta \mbF}W[\mcA_0,\mbF,da]_{\rm inv},
\label{eq:Jcov}
\end{align}
 where $W_{\rm inv}$ is now considered a functional of $\mcA_0$, $\mbF$, and $da$. 
 
 The anomalous consistent current can be obtained by varying $W_{\rm anom}$
\begin{align}
\mcJ_{0,\rm cons}&=T_0\frac{\delta }{\delta \mcA_0}W[\mcA_0,\mbA,da]_{\rm anom}, \nonumber \\[0.2cm] 
\mbJ_{\rm cons}&=T_0\frac{\delta }{\delta \mbA}W[\mcA_0,\mbA,da]_{\rm anom},
\end{align}
where $W_{\rm anom}$ has also to be considered as a functional of $\mcA_0$, $\mbA$, and $da$. Although the current $ \mbJ_{\rm cov}$ could also be computed by adding $\mcJ_{\rm BZ}$ to $\mcJ_{\rm cons}$, it is far simpler to obtain it directly from Eq.~(\ref{eq:Jcov}).
Since  only the covariant currents are relevant in hydrodynamics, we will give explicit expressions for them in four dimensions.
Using~\eqref{eq:Jcov} with~\eqref{eq:Winv4D} yields  
\begin{align}
\mcJ_{0,\rm cov}&=0, \nonumber \\[0.2cm]
\mbJ_{\rm cov}&=-{i\over 8\pi^{2}}\Big(\mcA_0\mbF+\mbF\mcA_0+\mcA_0^2da\Big).
\label{eq:Jcov4}
\end{align}
As usual, this formula directly gives $\mbJ_R(\mcA_R)_{\rm cov}$, as well as $\mbJ_L(\mcA_L)_{\rm cov}$ with a relative minus sign.

The fact that the covariant currents can be computed directly from $W_{\rm inv}$, which is obtained from the anomaly polynomial and is completely independent of the particular Chern-Simons form being used, has an important consequence: in a theory with left and right gauge fields, $(\mcA_R,\mcA_L)$ or their combinations $(\mcV,\mcA)$, the covariant currents are the same
independently of whether the anomalies are given in the left-right symmetric or the Bardeen form. 
Thus, there are two ways to compute the covariant currents in terms of vector and axial gauge fields, either from the expressions for $\mbJ_{R,\rm cov},\mbJ_{L,\rm cov}$ given by Eq.~\eqref{eq:Jcov4}
\begin{align}
\mbJ_V(\mcV,\mcA)_{\rm cov}&=\mbJ_R(\mcV+\mcA)_{\rm cov}+\mbJ_L(\mcV-\mcA)_{\rm cov}, \nonumber\\[0.2cm]
\mbJ_A(\mcV,\mcA)_{\rm cov}&=\mbJ_R(\mcV+\mcA)_{\rm cov}-\mbJ_L(\mcV-\mcA)_{\rm cov},
\end{align}
or directly from \eqref{eq:WinvVA4}, using 
\begin{align}
\mbJ_{V,\rm cov}&=T_0\frac{\delta }{\delta \mbF_V}W[\mcV_0,\mcA_0,\mbF_V,\mbF_A,da]_{\rm inv}, \nonumber \\[0.2cm] 
\mbJ_{A,\rm cov}&=T_0\frac{\delta }{\delta \mbF_A}W[\mcV_0,\mcA_0,\mbF_V,\mbF_A,da]_{\rm inv}.
\label{eq:JcovVA}
\end{align}
Either way, the final result in four dimensions is
\begin{align}
\mbJ_{V,\rm cov}&=-{i\over 4\pi^{2}}\Big[\mcV_0\mbF_A + \mbF_A\mcV_0+\mcA_0\mbF_V + \mbF_V\mcA_0+da(\mcV_0\mcA_0+\mcA_0\mcV_0 )
\Big],\nonumber\\[0.2cm]
\mbJ_{A,\rm cov}&=-{i\over 4\pi^{2}}\Big[\mcV_0\mbF_V + \mbF_V\mcV_0+\mcA_0\mbF_A + \mbF_A\mcA_0+da(\mcV_0^2+\mcA_0^2 )
\Big]\,.
\label{eq:JcovVA2}
\end{align}
Note that, as mentioned above, Eq.~\eqref{eq:Jbulk} implies that neither current is conserved.

\subsection{The energy-momentum tensor}
\label{sec:emtensor}

The Chern-Simons action (\ref{eq:CS_eff_action}) is topological, i.e independent of the gravitational background, so it cannot induce an anomalous 
energy-momentum tensor. In particular, there is no Bardeen-Zumino energy-momentum  tensor induced by $\Gamma[\mcA]_{\rm CS}$ on $\mcM_{2n-2}=\partial \mcM_{2n-1}$. On the other hand, we have seen that 
upon dimensional reduction on a stationary background,  
both the invariant and anomalous part of the partition function
$W_{\rm inv}$ and $W_{\rm anom}$ pick up a dependence on the metric functions 
$a_i$, which are conjugate to the  Kaluza-Klein invariant components $T_0^{\ i}$ of the energy-momentum tensor (see~\cite{Banerjee:2012iz} and Appendix \ref{app:formulae} for details). These are the only nonvanishing components of the
energy-momentum tensor, as the partition function is independent of 
the other metric functions 
$\sigma$ and $g_{ij}$. We denote them by its dual $(2n-4)$-form $\mbT$.

Thus, in order to compute this anomalous energy-momentum tensor it is enough to take the variation of
the anomalous piece of the partition function
$W_{\rm anom}$ with respect to $a$ [see Eq.~\eqref{eq:explicit_form_Ti0_fder}]. 
However, as in the case of the covariant gauge currents, it is much simpler to extract it from the boundary contribution of the variation of $W_{\rm inv}$. The vanishing  of the Bardeen-Zumino energy-momentum tensor guarantees that both methods give the same answer. In looking for those terms in $W_{\rm inv}$ depending on $da$,
we should keep in mind that this functional depends explicitly on $da$, but also implicitly through $\mbF= d\mbA+\mbA^2=-da\mcA_0+\ldots$ Then the same argument leading to~\eqref{eq:Jcov} gives in this case
\begin{align}
\mbT=T_0\left[\frac{\delta }{\delta (da)}-\mcA_0\frac{\delta }{\delta \mbF}\right]
W[\mcA_0,\mbF,da]_{\rm inv}.
\label{eq:Tcov}
\end{align}
In four dimensions, using Eq.~\eqref{eq:Winv4D} we find 
\begin{align}
\mbT={i\over 24\pi^{2}}{\rm Tr\,}\Big(3 \mcA_0^2\mbF+da \mcA_0^3\Big),
\label{eq:TR}
\end{align}
from where we directly read $\mbT(\mcA_R)$, and $\mbT(\mcA_L)$ with a relative minus sign. For a theory with 
vector and axial gauge fields, the anomalous energy-momentum tensor can be computed either from 
\begin{align}
\mbT(\mcV,\mcA)=\mbT(\mcV+\mcA)-\mbT(\mcV-\mcA)\,,
\end{align}
or by varying $W_{\rm inv}$
\begin{align}
\mbT(\mcV,\mcA)=T_0\left[\frac{\delta }{\delta (da)}-\mcV_0\frac{\delta }{\delta \mbF_V}-\mcA_0\frac{\delta }{\delta \mbF_A}\right]W[\mcV_0,\mcA_0,\mbF_V,\mbF_A,da]_{\rm inv}.
\end{align}
Either way, the result in four dimensions is given by
\begin{align}
\mbT(\mcV,\mcA)={i\over 4\pi^{2}}{\rm Tr\,}\left[  \mbF_V(\mcV_0\mcA_0+ \mcA_0\mcV_0) +\mbF_A(\mcV_0^2+\mcA_0^2)+\frac{1}{3}da(\mcA_0^3+3\mcA_0\mcV_0^2)
\right].
\label{eq:stress}
\end{align}

Before closing this section, we would like to stress that the existence of a nonvanishing anomalous contribution to the energy-momentum tensor is a direct consequence of the requirement of KK invariance of the partition function $W_{\rm anom}$. It is only through 
the KK invariant decomposition~\eqref{eq:KKdecomp} that the field $a=a_i\,dx^i$ enters both $W_{\rm anom}$ and $W_{\rm inv}$. Using 
\begin{align}
\mcA=\mcA_0 dx^0+\mcA_i dx^i\equiv\mcA_0 dx^0 +  \bm{\mcA},
\end{align}
instead of~\eqref{eq:KKdecomp} would give $\mbT=0$, but at the price that then neither the partition function nor the gauge currents would be KK invariant.
The situation is different in systems with spontaneously broken symmetry where, as we will see in the next section,  it is possible to have a non-anomalous energy-momentum tensor while preserving KK invariance.


\subsection{Example: Vector response in the presence of chiral imbalance}
\label{sec:example_2fQCD}

In order to illustrate the techniques presented so far, we construct the currents and
energy-momentum tensor for a physically interesting theory: two-flavor QCD coupled to an external electromagnetic 
field in a nontrivial ($a_{j}\neq 0$) background. In this example both chiral magnetic and vortical effects can
take place. It is known that in the presence of non-Abelian charges, only chemical potentials associated with mutually 
commuting charges can be considered \cite{Haber:1981ts,Yamada:2006rx}. 
Accordingly, we take the external fields taking values on the Cartan subalgebra of U(2) generated by 
$t_{0}$ and $t_{3}$ [see Eq. \eqref{eq:generatorsU(2)}]. In particular, we consider the field configuration
\begin{align}
\mcV_{0}&=\mcV_{00}t_{0}+\mcV_{03}t_{3}, \nonumber \\[0.2cm]
\mbV&=V_{0}t_{0}+V_{3}t_{3}, \nonumber \\[0.2cm]
\mcA_{0}&=\mcA_{00}t_{0}, \label{eq:2fQCD_field_conf}\\[0.2cm]
\mbA&=0. \nonumber
\end{align}
Since all external fields lie on the Cartan subalgebra, the corresponding field strengths are particularly simple
\begin{align}
\mbF_{V}&\equiv d\mbV+\mbV^{2}+\mbA^{2}=t_{0}dV_{0}+t_{3}dV_{3}, \nonumber\\[0.2cm]
\mbF_{A}&\equiv d\mbA+\mbA\mbV+\mbV\mbA=0.
\label{eq:2fQCD_field_strengths}
\end{align} 
The axial chemical potential $\mu_{5}$, controlling chiral imbalance, is introduced through \eqref{eq:A=A-muu}, which
in components reads
\begin{align}
\mcA_{0}&=\mu_{5}\mathbb{1}=2\mu_{5}t_{0} \hspace*{0.5cm}\Longrightarrow \hspace*{0.5cm} \mcA_{00}=2\mu_{5}, \nonumber \\[0.2cm]
\mcA_{i}&\equiv A_{i}+\mcA_{0}a_{i}=2a_{i}\mu_{5}t_{0}.
\label{eq:axial_fields_2fQCD}
\end{align}
By assuming $\mu_5$ to be constant, we are just only interested in the
one-derivative terms from the vectorial part of the background. 

The physical (i.e., KK invariant) electromagnetic gauge field $\mathbb{V}_{\mu}$
is related to the vector gauge field by the identities
\begin{align}
\label{eq:bgv}
\mcV_{0}&=eQ\mathbb{V}_{0}, \nonumber \\[0.2cm]
V_{i}&\equiv \mcV_{i}-\mcV_{0}a_{i}=eQ\mathbb{V}_{i}.
\end{align}
Here $eQ$ is the charge matrix for the two light flavors with
\begin{align}
Q=\left( 
\begin{array}{cc}
{2\over 3} &  0 \\
0 &  -{1\over 3} 
\end{array} 
\right) 
={1\over 3}t_{0}+t_{3}.
\end{align}
Using the expression of the electromagnetic and isospin currents 
\begin{align}
J_{\rm em}^{\mu}&=e \overline{\Psi}\gamma^{\mu}Q\Psi, \nonumber \\[0.2cm]
J_{\rm iso}^{\mu}&=\overline{\Psi}\gamma^{\mu}t_{3}\Psi,
\end{align} 
we find that the equilibrium currents can be expressed as   
the following combinations of the covariant currents
\begin{align}
\langle J^{\mu}_{\rm em}\rangle&={e \over 3}\langle J^{\mu}_{0,\rm cov}\rangle+e\langle J_{3,\rm cov}^{\mu}\rangle, \nonumber \\[0.2cm]
\langle J^{\mu}_{\rm iso}\rangle&=\langle J_{3,\rm cov}^{\mu}\rangle,
\end{align}
where the current expectation values $\langle J_{a,\rm cov}^{\mu}\rangle$ are given by the first equation in \eqref{eq:JcovVA2} 
by making the replacements 
\begin{align}
(\mcV_{00},V_{0i})&=  {e \over 3} (\mathbb{V}_{0},\mathbb{V}_{i}), \nonumber \\[0.2cm]
(\mcV_{03},V_{3i})&=  e(\mathbb{V}_{0},\mathbb{V}_{i}).
\end{align} 
Using in addition the expressions for the axial-vector fields in terms of the chemical potential given in Eq. \eqref{eq:axial_fields_2fQCD}, we
find
\begin{align}
\langle J^{i}_{\rm em} \rangle &= -{5e^{2}N_{c}  \over 18 \pi^2} \mu_5  \epsilon^{i j k}  \Big( 
  \partial_{j} \mathbb{V}_{k} + \mathbb{V}_{0} \partial_j a_k \Big), \nonumber \\[0.2cm]
\langle J^{i}_{\rm iso} \rangle &=-{eN_{c}\over 4 \pi^2} \mu_5 \epsilon^{ijk}  \Big( 
  \partial_{j} \mathbb{V}_{k} + \mathbb{V}_{0} \partial_{j} a_{k} \Big),
\label{eq:Js_2fQCD}
\end{align}
where $N_{c}$ is the number of colors.  
Inspecting these expressions we find contributions from both the chiral magnetic and chiral vortical effects. The first one is
associated with the term proportional to the magnetic field $B^{i}=\epsilon^{ijk}\partial_{j}\mathbb{V}_{k}$. 
Thus, from the first equation in \eqref{eq:Js_2fQCD}, we read the chiral magnetic conductivity \cite{Fukushima:2008xe,Kharzeev:2015znc}
\begin{align}
\sigma_{5}={e^2N_{c}{\rm Tr\,}Q^2\over 2\pi^{2}}\mu_{5}  ={5 e^2N_{c}\over 18 \pi^2}\mu_{5}.
\end{align}   
The vortical contribution, on the other hand, is identified as 
the term proportional to the curl of the KK field, $\epsilon^{ijk}\partial_{j}a_{k}$, associated
with the vorticity field, as shown in Eq.~\eqref{eq:fluid_variables2}. 

Similarly, we can use Eq.~\eqref{eq:stress} with the same replacements as above 
to write the anomaly-induced energy-momentum tensor 
\begin{align}
\langle T_{0}^{\,\,\,i} \rangle &={(6 \mu_5^{2}+5e^{2}\mathbb{V}_{0}^{2})N_{c}\over 36\pi^{2}} 
\mu_{5} \epsilon^{ijk}\partial_{j}a_{k} +
{5 e^{2}N_{c}\over 18\pi^2}\mu_{5}\mathbb{V}_{0} \epsilon^{ijk}\partial_{j}\mathbb{V}_{k},  \nonumber \\[0.2cm]
 \langle T_{00}\rangle&=\langle T^{ij}\rangle=0  .
\end{align}
Again, the nonvanishing components receive contributions from the chiral magnetic and vortical effects that can be identified as described. 
To the best of our knowledge, these explicit expressions for the energy-momentum tensor in the non-Abelian theory 
have not been reported before in the literature. They generalize the result found in \cite{Landsteiner:2012kd} for the U(1) case.


\section{The Wess-Zumino-Witten partition function}
\label{sec:WZW}

So far we have assumed that the symmetries we have dealt with, albeit maybe anomalous, are
preserved by the vacuum. For physical applications, however, it is convenient to consider situations in which these symmetries are
spontaneously broken, either total or partially. This is the case, for example, of chiral flavor symmetry in QCD, broken down to its vector subgroup, 
the electromagnetic gauge symmetry in conventional superconductors or the U(1) global phase in superfluids. 
Whenever this happens, Goldstone modes appear which couple to the macroscopic external gauge
fields and contribute to the anomaly. 
In this section we study the construction of partition functions for anomalous fluids with spontaneously broken symmetries extending
the previous analysis to the 
Wess-Zumino-Witten (WZW) action.

\subsection{Goldstone modes and the WZW action}

The WZW action describes the anomaly-induced interactions between the external gauge field $\mcA$ and the Goldstone bosons $\xi^a$. It admits a very simple expression~\cite{Zumino:1983ew} in terms of the Chern-Simons action introduced in Eq. \eqref{eq:CS_eff_action}
\begin{align}
\Gamma[\mcA,g]_{\rm WZW}=\Gamma[\mcA]_{\rm CS}-\Gamma[\mcA_g]_{\rm CS},
\label{eq:WZW}
\end{align} 
where $\mcA_g$ is the gauge transformed of $\mcA$ by the gauge group element $g\equiv\exp[-i\xi^a t_a]$
\begin{align}
\mcA_{g}=g^{-1}\mcA g+g^{-1}dg.
\label{eq:finitegauge}
\end{align}
Under a gauge transformation $h=e^u$, the Goldstone fields transform non-linearly according to  
\begin{align}
g_h=h^{-1}g.
\end{align}
This transformation makes $\mcA_g$  gauge invariant, namely
\begin{align}
(\mcA_h)_{g_h}=\mcA_{g} \hspace*{1cm} \Longrightarrow \hspace*{1cm} \delta_u\mcA_g=0,
\end{align}
and, as a consequence, $\delta_{u}\Gamma[\mcA_{g}]=0$. Thus, 
the gauge variation of the WZW action immediately gives the consistent anomaly 
\begin{align}
\delta_u\Gamma[\mcA,g]_{\rm WZW}= 
\delta_u \Gamma[\mcA]_{\rm CS}= -\int\limits_ {\mcM_{2n-2} }{\rm Tr\,}\Big(uG[\mcA(x)]_{\rm cons}\Big).
\label{eq:WZW_gauge}
\end{align} 

The structure of the WZW action can be better understood by using the transformation property of
the Chern-Simons form $\omega_{2n-1}^0$ under finite gauge transformations~\cite{Zumino:1983ew}
\begin{align}
\omega_{2n-1}^0(\mcA_g,\mcF_{g})=\omega_{2n-1}^0(\mcA,\mcF)+\omega_{2n-1}^0(dgg^{-1},0)+d\alpha_{2n-2}(\mcA,\mcF,g).
\label{eq:finitetrans}
\end{align} 
Here, $\omega_{2n-1}^{0}(dgg^{-1},0)$ is a $(2n-1)$-form proportional to $(dgg^{-1})^{2n-1}$ whereas $\alpha_{2n-2}(\mcA,\mcF,g)$ is a 
$(2n-2)$-form whose explicit expressions are given in Appendix~\ref{app:formulae}.
Then, Eq.~\eqref{eq:WZW} can be rewritten as 
\begin{align}
\Gamma[\mcA,g]_{\rm WZW}&=c_n\int\limits_{\mcM_{2n-1}}\Big[\omega_{2n-1}^{0}(\mcA)-\omega_{2n-1}^0(\mcA_g)\Big] 
\nonumber \\
&=-c_n\int\limits_{\mcM_{2n-1}}\omega_{2n-1}^0(dgg^{-1})-c_n\int\limits_{\mcM_{2n-2}}\alpha_{2n-2}(\mcA,g),
\label{eq:WZWloc}
\end{align} 
where, for simplicity we omit the field strength forms from the arguments.
Note that the dependence of the WZW action on the gauge field is given in terms of an integral over 
the $(2n-2)$-dimensional physical spacetime. In other words, unlike the Chern-Simons action, 
the WZW action is local and polynomial in the gauge fields.  This is however not true for the dependence on the Goldstone modes, 
due to the presence of the higher dimensional integral in the second line of Eq. \eqref{eq:WZWloc}. Using Eqs. 
\eqref{eq:winding} and \eqref{eq:alpha}, together with the identity \eqref{eq:WZWloc} we find in the four-dimensional case
($n=3$)
\begin{align}
\Gamma[\mcA,g]_{\rm WZW}={i\over 240\pi^{2}}\int\limits_{\mcM_{5}}{\rm Tr\,}(dgg^{-1})^{5} 
&-{i\over 48\pi^{2}}\int\limits_{\mcM_{4}}{\rm Tr\,}\Bigg[(dgg^{-1})(\mcA \mcF  + \mcF \mcA-\mcA^3)  \nonumber\\
&-\frac{1}{2}(dgg^{-1})\mcA(dgg^{-1})\mcA-\mcA(dgg^{-1})^3\Bigg].
\label{eq:WZWloc4}
\end{align} 

As usual, the previous expression gives the WZW action for a right-handed gauge field $\mcA_R$ coupled to a set of 
Goldstone fields $g_R\equiv\exp[-i\xi_R^a t_a]$, and can be used also for left-handed fields, with a relative minus sign. 
When both types of fields are present, the total WZW action will  be given by
\begin{align}
\Gamma[\mcA_R,\mcA_L,g_R,g_L]_{\rm WZW}=\Gamma[\mcA_R,g_R]_{\rm WZW}-\Gamma[\mcA_L,g_L]_{\rm WZW}.
\end{align} 
This  description in terms of two sets of Goldstone modes is appropriate when  the symmetry group $\mcG\times \mcG$ is completely broken. For applications to hadronic fluids, we are more interested in the case $\mcG\times \mcG\to\mcG$, where the symmetry is broken down to the diagonal subgroup of vector gauge transformations. In that case we have to halve the number of Goldstone bosons, in  correspondence with  the broken axial generators.

This can be accomplished by using the Chern-Simons form $\widetilde\omega_{2n-1}^0$ given in Eq. \eqref{om5_vector}
which preserves vector gauge transformations, instead of the left-right symmetric choice 
\eqref{eq:CS_Sym}. Thus, we have 
\begin{align}
\Gamma[\mcA_{L,R},g_{L,R}]_{\rm WZW}&=c_{n}\int\limits_{\mcM_{2n-1}}
\Big[\widetilde\omega_{2n-1}^0(\mcA_R,\mcA_L)-T(g_R,g_L)\widetilde\omega_{2n-1}^0(\mcA_R,\mcA_L)\Big]
\nonumber\\[0.2cm]
&\equiv \widetilde\Gamma[\mcA_R,\mcA_L]_{\rm CS}-\widetilde\Gamma[\mcA_R^{g_R},\mcA_L^{g_L}]_{\rm CS},
\label{eq:eff_act_GB_1}
\end{align} 
where we have introduced the notation
\begin{align}
T(g)f(\mcA)\equiv f(\mcA_g).
\end{align} 
Now we have to take into account that the transformation \eqref{eq:finitegauge} implies the following 
group composition law 
\begin{align}
T(g_1)T(g_2)=T(g_2g_1).
\end{align}
Remembering also that $\widetilde\omega_{2n-1}^0$ is invariant under finite vector gauge transformations $T(g,g)$, 
we can write
\begin{align}
T(g_R,g_L)\widetilde\omega_{2n-1}^0(\mcA_R,\mcA_L)&=T(g_R,g_R)T(e,g_Lg_R^{-1})\widetilde\omega_{2n-1}^0(\mcA_R,\mcA_L)
\nonumber \\[0.2cm]
&=T(e,U)\widetilde\omega_{2n-1}^0(\mcA_R,\mcA_L)
\end{align}
where $e$ represents the identity and 
\begin{align}
U=g_Lg_R^{-1}\equiv e^{2i\xi^a t_a},
\end{align} 
is given in terms of a single set of Goldstone fields and 
transforms under $\mathcal{G}\times\mathcal{G}$ according to
\begin{align}
U\to h_L^{-1}Uh_R.
\end{align} 
With all these ingredients in mind, we find that the appropriate WZW action takes the form 
\begin{align}
\Gamma[\mcA_R,\mcA_L,U]_{\rm WZW}=\widetilde\Gamma[\mcA_R,\mcA_L]_{\rm CS}
-\widetilde\Gamma[\mcA_R,\mcA_L^U]_{\rm CS},
\label{eq:WZW_Bardeen}
\end{align}
where 
\begin{align}
\mcA_L^U=U^{-1}\mcA_L U+U^{-1}dU,
\end{align} 
and $\widetilde\Gamma[\mcA_R,\mcA_L]_{\rm CS}$ 
is the integral of $\widetilde\omega_{2n-1}^0(\mcA_R,\mcA_L)$ over $\mcM_{2n-1}$, as shown in \eqref{eq:eff_act_GB_1}. 
Note that, by construction, this definition of the WZW action gives the correct anomaly in the Bardeen form.

We can obtain a more useful expression for the WZW action by using the property
\begin{align}
\widetilde\omega_{2n-1}^0(\mcA_R,\mcA_L)=\omega_{2n-1}^0(\mcA_R,\mcA_L)+dS_{2n-2}(\mcA_R,\mcA_L)
\label{eq:S_2n}
\end{align}
where $S_{2n-2}$ is the Bardeen counterterm (see Appendix~\ref{app:formulae} for explicit formulae). 
Using 
(\ref{eq:WZW_Bardeen}) together with (\ref{eq:finitetrans}) and (\ref{eq:S_2n}) finally gives
\begin{align}
\Gamma[\mcA_R,\mcA_L,U]_{\rm WZW}&=c_n\int\limits_{\mcM_{2n-1}}\omega_{2n-1}^0(dUU^{-1})\nonumber\\
+&\,c_n\int\limits_{\mcM_{2n-2}}\Big[\alpha_{2n-2}(\mcA_L,U)+S_{2n-2}(\mcA_R,\mcA_L)-S_{2n-2}(\mcA_R,\mcA_L^U)\Big].
\end{align}
The explicit expression of $\Gamma_{\rm WZW}$ in four dimensions is rather long and will not be given here. 
It can be found, for instance, in Refs.~\cite{Witten:1983tw,Kawai:1984mx,Manes:1984gk}.


\subsection{Gauge currents in general backgrounds}

The formula for $\Gamma_{\rm WZW}$ given in \eqref{eq:WZWloc} shows that, unlike the Chern-Simons action, the WZW effective action does \emph{not} induce any bulk current. Instead,  a general variation $\delta\mcA=B$  induces only a local consistent gauge current  on the boundary spacetime $\mcM_{2n-2}$
\begin{align}
\delta_B\Gamma[\mcA,g]_{\rm WZW}&=-c_n\delta_B\int\limits_{\mcM_{2n-2}}\alpha_{2n-2}(\mcA,g) \nonumber \\
&=\int\limits_{\mcM_{2n-2}} {\rm Tr\,}\Big[B\mcJ(\mcA,g)_{\rm cons}\Big].
\end{align} 
This is a consistent current, as follows from the fact that adding to it the Bardeen-Zumino current yields a covariant current. Indeed, combining~\eqref{eq:WZW} with~\eqref{eq:CS_currents} gives
\begin{align}
\delta_B\Gamma[\mcA,g]_{\rm WZW}&=\delta_{B}\Gamma[\mcA]_{\rm CS}-\delta_{B}\Gamma[\mcA_g]_{\rm CS}\nonumber\\[.2cm]
&=\int\limits_{\mcM_{2n-1}} \,\Big\{ {\rm Tr\,}\Big[B\mcJ(\mcA)_{\rm bulk}\Big]- {\rm Tr\,}\Big[B_g\mcJ(\mcA_g)_{\rm bulk}\Big]
\Big\}\nonumber\\
&-\int\limits_{\mcM_{2n-2}} \,\Big\{ {\rm Tr\,}\Big[B\mcJ(\mcA)_{\rm BZ}\Big]- {\rm Tr\,}\Big[B_g\mcJ(\mcA_g)_{\rm BZ}\Big]
\Big\}
\label{eq:delta_WZW}.
\end{align} 
Then, using that both $B$ and $\mcJ(\mcA)_{\rm bulk}$ transform covarianly
\begin{align}
B_g&=g^{-1}Bg, \nonumber \\[0.2cm] \mcJ(\mcA_g )_{\rm bulk} &= g^{-1}\mcJ(\mcA)_{\rm bulk}\,g,
\end{align} 
we see that the first integral in the right-hand side of~\eqref{eq:delta_WZW} vanishes, yielding
\begin{align}
\delta_B\Gamma[\mcA,g]_{\rm WZW}=\int\limits_{\mcM_{2n-2}} \, {\rm Tr\,}\Big\{B\Big[g\mcJ(\mcA_g)_{\rm BZ}g^{-1}
-\mcJ(\mcA)_{\rm BZ}\Big]\Big\}\,.
\end{align}
Given  the gauge invariance of $\mcA_g$, the first term in the right-hand side can be identified with the covariant gauge current, and we conclude that
\begin{align}
\mcJ(\mcA,g)_{\rm cons}&=g\mcJ(\mcA_g)_{\rm BZ}g^{-1}-\mcJ(\mcA)_{\rm BZ},\nonumber\\[0.1cm]
\mcJ(\mcA,g)_{\rm cov}&=g\mcJ(\mcA_g)_{\rm BZ}g^{-1}=\mcJ(\mcA+dgg^{-1})_{\rm BZ}.
\end{align}

This remarkable connection between the  gauge currents and the Bardeen-Zumino current  provides the most efficient computational method in the presence of  spontaneous symmetry breaking, bypassing the need to use the WZW action \cite{Lublinsky:2009wr,Lin:2011mr,Nair:2011mk,Fukushima:2012fg}. Indeed, using \eqref{eq:BZ4} we easily find the covariant current in four dimensions
\begin{align}
\mcJ(\mcA,g)_{\rm cov}={i\over 24\pi^{2}}\left[\mcF(\mcA+dgg^{-1})+(\mcA+dgg^{-1})\mcF-\frac{1}{2}(\mcA+dgg^{-1})^3\right].
\label{eq:JcovSSB4}
\end{align}
This is valid for a right-handed fermion, while as usual the left-handed current differs just by a relative minus sign. 

This can be easily extended to the phenomenologically interesting case  $\mcG\times \mcG\to\mcG$, where the symmetry is broken down to the diagonal subgroup. Basically, one just have to make the replacements $\mcA\to(\mcA_R,\mcA_L)$ and $g\to(e,U)$ in the previous relations, obtaining
\begin{align}
\mcJ^R(\mcA_R,\mcA_L,U)_{\rm cov}&=\mcJ^R(\mcA_R,\mcA_L^U)_{\rm BZ} \nonumber\\[0.2cm]
\mcJ^L(\mcA_R,\mcA_L,U)_{\rm cov}&=U\mcJ^L(\mcA_R,\mcA_L^U)_{\rm BZ}U^{-1}\,,
\label{eq:JcovWZW}
\end{align}
where the Bardeen-Zumino currents can be computed from the relation [cf. \eqref{eq:BZhom}]
\begin{align}
c_n \ell\,\widetilde\omega_{2n-1}^{0}(\mcA_R,\mcA_L)=-{\rm Tr\,}\Big(B_R\mcJ^{R}_{\rm BZ}+B_L\mcJ^{L}_{\rm BZ}\Big).
\end{align}

In four dimensions, the Bardeen-Zumino currents read
\begin{align}
\mcJ^{R}_{\rm BZ}(\mcA_{L},\mcA_{R})&=-{i\over 24\pi^{2}}\left[(\mcA_L-\mcA_R)\left(\mcF_R+\frac{1}{2}\mcF_L\right) \right.\nonumber\\[0.2cm]
&+\left.\left(\mcF_R+\frac{1}{2}\mcF_L\right)(\mcA_L-\mcA_R)-\frac{1}{2}(\mcA_L-\mcA_R)^3\right], \nonumber\\[0.2cm]
\mcJ^{L}_{\rm BZ}(\mcA_{L},\mcA_{R})&=-{i\over 24\pi^{2}}\left[(\mcA_L-\mcA_R)\left(\mcF_L+\frac{1}{2}\mcF_R\right) \right. \\[0.2cm]
&+\left.\left(\mcF_L+\frac{1}{2}\mcF_R\right)(\mcA_L-\mcA_R)-\frac{1}{2}(\mcA_L-\mcA_R)^3\right]. 
\nonumber
\end{align}
Now, using \eqref{eq:JcovWZW} we can obtain the following expressions for the covariant current in the presence of Goldstone modes
\begin{align}
\mcJ^R(\mcA_R,&\mcA_L,U)_{\rm cov}=-{i\over 24\pi^{2}}\left[\Big(U^{-1}\mcA_LU-\mcA_R+U_R\Big)\left(\mcF_R+\frac{1}{2}U^{-1}\mcF_LU
\right)\right. \nonumber\\[0.2cm]
&+\left.\left(\mcF_R+\frac{1}{2}U^{-1}\mcF_LU\right)\Big(U^{-1}\mcA_LU-\mcA_R+U_R\Big)
-\frac{1}{2}\Big(U^{-1}\mcA_LU-\mcA_R+U_R\Big)^3\right], \nonumber \\[0.2cm]
\mcJ^L(\mcA_R,&\mcA_L,U)_{\rm cov}=-{i\over 24\pi^{2}}\left[\Big(\mcA_L-U\mcA_R\,U^{-1}+U_L\Big)\left(\mcF_L+\frac{1}{2}U\mcF_RU^{-1}\right)\right. 
\label{eq:JLcov_JRcov}\\[0.2cm] &+\left.\left(\mcF_L+\frac{1}{2}U\mcF_RU^{-1}\right)\Big(\mcA_L-U\mcA_R\,U^{-1}+U_L\Big)
-\frac{1}{2}\Big(\mcA_L-U\mcA_R\,U^{-1}+U_L\Big)^3\right],
\nonumber
\end{align}
where we have defined the adjoint fields
\begin{align}
U_R&\equiv U^{-1}dU, \nonumber \\[0.2cm] 
U_L&\equiv dUU^{-1}.
\label{eq:ULUR_def}
\end{align}
From these results, we can obtain as well the vector and axial gauge currents as functions of $\mcV$ and $\mcA$, using
\begin{align}
\mcJ^V(\mcV,\mcA,U)_{\rm cov}&=\mcJ^R(\mcV+\mcA,\mcV-\mcA,U)_{\rm cov}+\mcJ^L(\mcV+\mcA,\mcV-\mcA,U)_{\rm cov},
\nonumber\\[0.2cm]
\mcJ^A(\mcV,\mcA,U)_{\rm cov}&=\mcJ^R(\mcV+\mcA,\mcV-\mcA,U)_{\rm cov}-\mcJ^L(\mcV+\mcA,\mcV-\mcA,U)_{\rm cov},
\end{align}
but the explicit expressions resulting from the substitution of Eq. \eqref{eq:JLcov_JRcov} are rather cumbersome.


\subsection{Stationary backgrounds and the WZW partition function}
The partition function with spontaneously broken symmetry can be readily obtained  from the corresponding one in
the absence of Goldstone bosons by applying \eqref{eq:WZW} in a time-independent background
\begin{align}
W[\mcA_0,\mbA,g]_{\rm WZW}=i\Gamma[\mcA_0,\mbA,g]_{\rm WZW}=i\Gamma[\mcA_0,\mbA]_{\rm CS}-i\Gamma[\mcA_{0g},\mbA_g]_{\rm CS}.
\end{align}
Then, using~\eqref{eq:dired}, we find that the invariant part cancels and the partition function can be written only
in terms of the local anomalous part as
\begin{align}
W[\mcA_0,\mbA,g]_{\rm WZW}=W[\mcA_0,\mbA]_{\rm anom}-W[\mcA_{0g},\mbA_g]_{\rm anom},
\label{eq:WZWpart}
\end{align}
where, to avoid a cumbersome notation here and in the following, we have omitted the dependence on the metric function $a_{i}$.
Noting  that $\mcA_0$ transforms covariantly as an adjoint field under time-independent gauge transformations, $\mcA_{0g}=g^{-1}\mcA_0 g$, and using the cyclic property of the trace, the formula for the partition function can be 
simplified to
\begin{align}
W[\mcA_0,\mbA,g]_{\rm WZW}=W[\mcA_0,\mbA]_{\rm anom}-W[\mcA_0,\mbA+dgg^{-1}]_{\rm anom}.
\end{align} 
Using the explicit form of $W[\mcA_0,\mbA]_{\rm anom}$ given in \eqref{eq:Wanom4}, yields the partition function in four dimensions
\begin{align}
W[\mcA_0,\mbA,g]_{\rm WZW}&={i\over 24\pi^{2}T_{0}}\int_{S^{3}} {\rm Tr\,}\left\{\mcA_0\left[\mbF dgg^{-1}+dgg^{-1}\mbF+\frac{1}{2}\mbA^3 \right.\right.\nonumber \\[0.2cm]
&-\left.\left.\frac{1}{2}\Big(\mbA+dgg^{-1}\Big)^3\right] +\mcA_0^2\,dgg^{-1}da\right\}.
\end{align}

The result \eqref{eq:WZWpart} can be obviously extended to the case where the gauge group is spontaneously broken to its
diagonal vector subgroup, $\mcG\times \mcG\to\mcG$. In this case, using the partition function in the Bardeen form, the WZW action reads
\begin{align}
W[\mcV_0,\mcA_0,\mbV,\mbA,U]_{\rm WZW}=W[\mcV_0,\mcA_0,\mbV,\mbA]_{\rm anom}-W[\mcV_0^U,\mcA_0^U,\mbV^U,\mbA^U]_{\rm anom},
\label{eq:WZW_VA_basis_prescription}
\end{align}
where $\mcV_0$ and  $\mcA_0$ transform under time-independent gauge transformations as
\begin{align}
\mcV_0^U&=\frac{1}{2}(\mcV_0+\mcA_0)+\frac{1}{2}U^{-1}(\mcV_0-\mcA_0)U,\nonumber\\[0.2cm]
\mcA_0^U&=\frac{1}{2}(\mcV_0+\mcA_0)-\frac{1}{2}U^{-1}(\mcV_0-\mcA_0)U,
\label{eq:subs1}
\end{align}
with identical transformations for $\mbF_V$ and $\mbF_A$. On the other hand, the transformation of $\mbV$ and $\mbA$ 
picks up an extra non-homogeneous term
\begin{align}
\mbV^U&=\frac{1}{2}(\mbV+\mbA)+\frac{1}{2}U^{-1}(\mbV-\mbA)U+\frac{1}{2}U^{-1} dU, \nonumber\\[0.2cm]
\mbA^U&=\frac{1}{2}(\mbV+\mbA)-\frac{1}{2}U^{-1}(\mbV-\mbA)U-\frac{1}{2}U^{-1} dU.
\label{eq:subs2}
\end{align}
Using the transformation laws \eqref{eq:subs1} and \eqref{eq:subs2} on \eqref{eq:WanomVA4} yields the partition function\break \hbox{$W[\mcV_0,\mcA_0,\mbV,\mbA,U]_{\rm WZW}$}. Although the computation is straightforward the result is rather long and will not be reproduced here.

To illustrate this technique, we consider again the exampled discussed in Sec. \ref{sec:example_2fQCD},
two-flavor QCD in four dimensions, where
now the global symmetry group U(2)$_{L}\times$U(2)$_{R}$ is broken down to its vector subgroup U(2)$_{V}$.
The WZW action can be computed using the prescription
\eqref{eq:WZW_VA_basis_prescription} with the anomalous piece of the effective action obtained in Eq. \eqref{eq:W_anom_VA_U2xU2}. In doing this,
we have to keep in mind that the U(1)$_{A}$ factor is broken at the nonperturbative level, so there is no Goldstone mode associated with 
U(1)$_{A}$ and the field $U$ only has components
on the SU(2)$_{A}$ factor. 
Again, the external fields are taken to lie in the Cartan subalgebra in the field configuration defined 
in Eq. \eqref{eq:2fQCD_field_conf}, with the corresponding field strengths given in Eq. \eqref{eq:2fQCD_field_strengths}.
Using these expressions, we find from
\eqref{eq:W_anom_VA_U2xU2} that $W[\mcV_{0},\mcA_{0},\mbV,\mbA]_{\rm anom}=0$ for $\mbA=0$. 
On the other hand, to compute the second term in \eqref{eq:WZW_VA_basis_prescription}, we need Eq. \eqref{eq:subs1}, together with
\begin{align}
\mbA^{U}&={1\over 2}V_{3}\Big(t_{3}-U^{-1}t_{3}U\Big)-{1\over 2}U^{-1}dU, \nonumber \\[0.2cm]
\mbF_{V}^{U}&=t_{0}dV_{0}+{1\over 2}dV_{3}\Big(t_{3}+U^{-1}t_{3}U\Big), \\[0.2cm]
\mbF_{A}^{U}&={1\over 2}dV_{3}\Big(t_{3}-U^{-1}t_{3}U\Big).
\nonumber
\end{align}
Notice that, from the first and last equations, we find 
\begin{align}
{\rm Tr\,}\mbA^{U}&=-{1\over 2}{\rm Tr\,}\Big(U^{-1}dU\Big)=0, \nonumber \\[0.2cm]
{\rm Tr\,}\mbF^{U}_{A}&=0,
\end{align}
since $U$ only takes value on SU(2). We have then
\begin{align}
W[\mcV_0,\mcA_0,\mbV,\mbA,U]_{\rm WZW}^{{\rm U}(2)\times{\rm U}(2)}
&={i\over 4\pi^{2} T_0}\int\limits_{S^{3}} \left\{   
\big({\rm Tr\,}\mcV_0^{U}\big){\rm Tr\,}\left[\mbF_V^{U} \mbA^{U}-{2\over 3}\big(\mbA^{U}\big)^3\right]
\right.\nonumber \\[0.2cm]
&+\big({\rm Tr\,}\mbF_{V}^{U}\big){\rm Tr\,}\big(\mcV_{0}^{U}\mbA^{U}\big)
+{1\over 3}\big({\rm Tr\,}\mcA_{0}^{U}\big){\rm Tr\,}\big(\mbF_{A}^{U}\mbA^{U}\big)
\\[0.2cm]
&\left.+da\big({\rm Tr\,}\mcV_{0}^{U}\big){\rm Tr\,}\big(\mcV_{0}^{U}\mbA^{U}\big) 
+{1\over 3}da\big({\rm Tr\,}\mcA_{0}^{U}\big){\rm Tr\,}\big(\mcA_{0}^{U}\mbA^{U}\big) \right\},
\nonumber
\end{align}
which upon substitution renders the result
\begin{align}
W[\mcV_0,\mcA_0,\mbV,\mbA,U]_{\rm WZW}^{{\rm U}(2)\times{\rm U}(2)}
&={i\over 8\pi^{2} T_0}\int\limits_{S^{3}} \left\{ 
{1\over 2}\mcV_{00}V_{3}{\rm Tr\,}\Big[t_{3}d(U_{R}+U_{L})\Big]  
+{1\over 6}\mcV_{00}{\rm Tr\,}\big(U_{L}^{3}\big)
\right. \nonumber \\[0.2cm]
&-{1\over 2}\Big(\mcV_{00}dV_{3}+\mcV_{03}dV_{0}+da\mcV_{00}\mcV_{03}\Big){\rm Tr\,}\Big[t_{3}\big(U_{R}+U_{L}\big)\Big]
\label{eq:general_SSB_U2xU2} \\[0.2cm]
&-{1\over 6}\mcA_{00}\Big(dV_{3}+da\mcV_{03}\Big){\rm Tr\,}\Big[t_{3}(U_{R}-U_{L})\Big] \nonumber \\[0.2cm]
&\left.+{1\over 3}\mcA_{00}\Big(dV_{3}V_{3}+da \mcV_{03} V_{3}\Big){\rm Tr\,}\Big[t_{3}\Big(t_{3}-U^{-1}t_{3}U\Big)\Big]\right\},
\nonumber
\end{align}
where we have used $U_{R}$, $U_{L}$ defined in Eq. \eqref{eq:ULUR_def}.
This equation is one of the main results of our work and illustrates the power of the differential geometry methods introduced. 

It is instructive to compute the previous partition function for the particular case of the chiral-imbalanced electromagnetic background 
studied in Sec. \ref{sec:example_2fQCD}, this time on a flat background $(a_{k}=0)$. Here we just give the results, the full calculation being deferred to a future work \cite{in_progress}.
Using the Pauli matrices, the unitary Goldstone matrix $U$ can be parametrized as   
\begin{align}
U = \exp\left(i\sum_{a=1}^{3}\zeta_{a}\sigma_{a}\right), 
\end{align}
where, in terms of the conventionally normalized pion fields $\{\pi^{0}$, $\pi^{\pm}\}$, we write
\begin{align}
\sum_{a=1}^{3} \zeta_{a} \sigma_{a}={\sqrt{2}\over f_\pi} 
\left(
\begin{array}{cc}
\tfrac{1}{\sqrt{2}} \pi^0 & \pi^+ \\
\pi^- & -\tfrac{1}{\sqrt{2}} \pi^0
\end{array} \right),  
\end{align}
with $f_\pi\approx 92$ MeV the pion decay constant.  
Expanding the effective action~\eqref{eq:general_SSB_U2xU2} in powers of the pion fields, after a lengthy calculation we find the following
expression for the partition function  
\begin{align}
T_0 W &=\int d^{3}x\left[{e^2 N_{c} \over 12 \pi^2 f_{\pi}}  \mathbb{V}_0 \, \partial_i \pi^0 B^i  
-{ie\mu_5  N_c \over 12 \pi^2 f_\pi^2} \Big(\pi^{-}\partial_{j}\pi^{+}-\pi^{+}\partial_{j}\pi^{-}  
-2ie\pi^{-}\pi^{+}\mathbb{V}_{j}\Big)B^{j} \right.
\nonumber \\[0.2cm]
&
+\mathcal{O}(\pi^{3})\bigg], 
\label{eq:W_weakapprox}
\end{align}
where $B^{i} = \epsilon^{ijk} \partial_{j} \mathbb{V}_{k}$ is the magnetic field. 
The first term on the right-hand side of Eq.~(\ref{eq:W_weakapprox}), linear in the pion field, is in perfect agreement with 
the magnitude of the term in the effective Lagrangian giving the electromagnetic decay of the neutral pion, $\pi^0 \to 2 \gamma$
\begin{equation}
\mathscr{L}_{\rm eff} \supset \frac{e^{2}N_{c}}{96\pi^{2}f_{\pi}}\pi^{0}\epsilon^{\mu\nu\alpha\beta}\mathbb{F}_{\mu\nu}\mathbb{F}_{\alpha\beta},
\end{equation} 
where $\mathbb{F}_{\mu\nu}=\partial_{\mu}\mathbb{V}_{\nu}-\partial_{\nu}\mathbb{V}_{\mu}$ is
the field strength associated with the electromagnetic potential~$\mathbb{V}_{\mu}$. 
It also reproduces the result found in Ref. \cite{Son:2007ny}.
In addition, the quadratic term in~(\ref{eq:W_weakapprox}) also agrees with the known form 
of the parity-odd couplings obtained from the Wess-Zumino-Witten action in the presence of chiral imbalance
(see, for example, Refs.~\cite{Witten:1983tw,Andrianov:2017ely}).

\subsection{Gauge currents and energy-momentum tensor in stationary backgrounds with spontaneous symmetry breaking}

The application of this formalism to hydrodynamics requires the knowledge of the covariant gauge currents in stationary backgrounds~\cite{Lin:2011aa}. 
One possible route is to compute the consistent currents from a general variation 
\eqref{eq:B_general_varA} of the partition function, which then can be covariantized through the addition of the (dimensionally reduced) Bardeen-Zumino current. A more efficient way is to perform the dimensional reduction of the covariant gauge currents, which in the presence of symmetry breaking admit explicit, local expressions
[cf. Eqs. (\ref{eq:JcovSSB4}) and (\ref{eq:JLcov_JRcov}) in four dimensions].

An efficient way to carry out the dimensional reduction of the currents is by writing
\begin{align}
\int\limits_{\mcM_{2n-2}}{\rm Tr\,}\Big[B \mcJ(\mcA,\mcF,g)\Big]=\frac{1}{T_0}\int\limits_{S^{2n-3}}{\rm Tr\,}\Big[\mcB_0 \mcJ_0(\mcA_0,\mbA,g)+\mbB \mbJ (\mcA_0,\mbA,g)\Big],
\label{eq:Bcurrents}
\end{align}
and using the identity
\begin{align}
\hspace*{-0.2cm}\int\limits_{\mcM_{2n-2}}f(B,\mcA,\mcF)=\frac{1}{T_0}\int\limits_{S^{2n-3}}\left.\left( 
\mcB_0\frac{\delta}{\delta B}+\mcA_0\frac{\delta}{\delta\mcA}-\mbD\mcA_0 \frac{\delta}{\delta\mcF}
 \right)              
f(B,\mcA,\mcF)\right|_{\substack{\hspace*{-0.2cm}\mcA,B\to\mbA,\mbB \\[0.1cm]\mcF\to\mbF+\mcA_0da}},
\label{eq:intf(BAF)}
\end{align}
where $\mbD$ is the covariant derivative with respect to the connection $\mbA$ defined in Eq. \eqref{eq:covariant_derivative_bfA}. 
This equation follows from the decompositions \eqref{eq:KKdecomp} and \eqref{eq:B_general_varA}, together with \eqref{eq:field_strenght_decomp_theta_da},
which reads
\begin{align}
\mcF=\mbF+\mcA_0da-\theta\mbD\mcA_0.
\label{eq:F0bF+A0da-thetaDA0}
\end{align}
Notice that this identity introduces terms proportional to $dx^0$ that survive upon dimensional reduction on the thermal cycle, hence the functional
derivative with respect to $\mcF$. Applying Eq.~\eqref{eq:intf(BAF)} to \eqref{eq:Bcurrents} immediately yields
\begin{align}
\mcJ_0(\mcA_0,\mbA,\mbF,g)=\mcJ(\mcA,\mcF,g)\Bigg|_{\substack{\hspace*{-1cm}\mcA\to\mbA \\[0.1cm]\mcF\to\mbF+\mcA_0da}},
\label{eq:J0dr}
\end{align}
for the anomalous non-abelian charge density, and
\begin{align}
\mbJ(\mcA_0,\mbA,\mbF,g)=\left.\left( 
-\mcA_0\frac{\delta}{\delta\mcA}+\mbD\mcA_0 \frac{\delta}{\delta\mcF}
 \right)              
\mcJ(\mcA,\mcF,g)\right|_{\substack{\hspace*{-1cm}\mcA\to\mbA \\[0.1cm]\mcF\to\mbF+\mcA_0da}},
\label{eq:Jdr}
\end{align}
for the anomalous current. 

We finally study the four-dimensional case. Using Eq.~\eqref{eq:JcovSSB4} for the covariant current gives
\begin{align}
\mcJ_0(\mcA_0,\mbA,\mbF,g)_{\rm cov}&={i\over 24\pi^{2}}\Bigg[(\mbF+\mcA_0da)(\mbA+dgg^{-1})
\nonumber \\[0.2cm]
&+(\mbA+dgg^{-1})(\mbF+\mcA_0da)-\frac{1}{2}(\mbA+dgg^{-1})^3\Bigg],\nonumber
\end{align}
and
\begin{align}
\mbJ(\mcA_0,\mbA,\mbF,g)_{\rm cov}&={i\over 24\pi^{2}}\Bigg[ \mbD\mcA_0\Big(\mbA+dgg^{-1}\Big)-\Big(\mbA+dgg^{-1}\Big)\mbD\mcA_0
\nonumber\\[0.2cm]
&- \Big(\mbF+\mcA_0da\Big)\mcA_0-\mcA_0\Big(\mbF+\mcA_0da\Big)+\frac{1}{2}\mcA_0\Big(\mbA+dgg^{-1}\Big)^2 
 \\[0.2cm]
&-\frac{1}{2}\Big(\mbA+dgg^{-1}\Big)\mcA_0\Big(\mbA+dgg^{-1}\Big)+\frac{1}{2}\Big(\mbA+dgg^{-1}\Big)^2\mcA_0\Bigg].   
\nonumber
\end{align}
As in many other instances along this paper, it should be clear that Eqs. \eqref{eq:J0dr} and \eqref{eq:Jdr} can also be applied to a system with left and right (or vector and axial-vector) gauge currents, with obvious modifications.

Finally, it is easy to see that the anomalous energy-momentum tensor must vanish in a system with spontaneously broken symmetry described by the WZW action. The reason is that, as can be seen from Eq. \eqref{eq:Tcov}, the anomalous components $T_0^{\ i}$ of the energy-momentum tensor in stationary backgrounds are  invariant 
under time-independent gauge transformations
\begin{align}
T_0^{\ i}(\mcA_0^g,\mbA_g)=T_0^{\ i}(\mcA_0,\mbA).
\end{align} 
Thus, Eq. \eqref{eq:WZWpart}
implies that the anomalous energy-momentum tensor is zero in the presence of spontaneous symmetry breaking. Indeed, computing its only 
potentially nonvanishing component, we find that the contribution of the gauge fields is exactly cancelled by that of the Goldstone modes
\begin{align}
T_0^{\ i}(\mcA_0,\mbA,g)=T_0^{\ i}(\mcA_0,\mbA) - T_0^{\ i}(\mcA_0^g,\mbA_g)=0.
\end{align}


\section{Discussion and outlook}
\label{sec:discussion}

Differential geometry is indeed a powerful tool to address the issue of quantum field theory anomalies. 
In this work we have employed it to carry out a systematic construction of partition functions for nondissipative fluids in the 
presence of non-Abelian anomalies. We have provided explicit expressions for the anomalous contribution to
the fluid partition function for generic theories in terms of
functional derivatives of the corresponding Chern-Simons effective action. More importantly, our analysis can be also applied 
to theories with spontaneous symmetry breaking, in which
case the starting point is the WZW effective action for Goldstone modes, which is straightforwardly constructed in terms of the 
corresponding partition functions. 

We have also studied in detail the gauge currents and energy-momentum tensor induced by the anomaly, giving operational 
expressions for these quantities. The covariant current, being determined solely from the nonlocal invariant piece of the 
partition function \cite{Jensen:2013kka},
is therefore independent of any local counterterms modifying the anomalous part. This in particular means that its form does not
depend on using either the symmetric or the Bardeen form of the anomaly.
In the case of the energy-momentum tensor, we have shown that the requirement of KK invariance of the low energy fields 
forces a nonvanishing anomaly-induced contribution. The situation is quite different for theories with spontaneous symmetry breaking.
In this case, the Bardeen-Zumino current fully determines both the consistent and covariant currents. These can be 
computed either from the variation of the partition function or, alternatively, 
from the dimensional reduction of the Bardeen-Zumino current. In the case of the induced energy-momentum
tensor, we find a vanishing result due to the cancellation between the contribution of gauge fields and Goldstone modes.

There are a number of issues that can be effectively addressed using the methods described in this work. 
For example, the anomaly-induced equilibrium partition functions obtained here only include the effect of 
chiral anomalies. This is clear from the fact that our effective actions only contain first derivatives of the
metric functions, whereas gravitational anomalies depend on the curvature two-form which includes second-order derivatives. A 
next step is  
to incorporate the effect of gravitational and mixed gauge-gravitational 
anomalies into the formalism, thus generalizing existing analysis in the literature (e.g.,~\cite{Jensen:2012kj,Jensen:2013kka}). 

This can be done following the same strategy used in this work for chiral anomalies: implementing dimensional reduction on the Chern-Simons
form associated with the appropriate anomaly polynomial, which now includes both the contributions of the gauge fields and the background 
curvature~\cite{AlvarezGaume:1983ig,AlvarezGaume:1984dr}. 
Using homotopy techniques
similar to the ones employed in this paper, it
is possible to write general formulae for the Chern-Simons form in any dimension, including also the contribution of the 
background curvature. This opens the way to give general prescriptions
for the construction of equilibrium partition functions for fluids including the effects of gravitational anomalies. This problem
will be addressed in detail elsewhere.

Finally, in this article we have confined our attention to the application of differential geometry methods to give a general prescription for the construction of nondissipative partition functions. These can be applied to the study of a variety of hydrodynamic systems with non-Abelian anomalies. 
A natural task now is to exploit these techniques to study the constitutive relations, as well as the corresponding anomalous 
transport coefficients, of a variety of systems of
physical interest, including hadronic fluids relevant for the physics of heavy ion 
collisions~\cite{Yin:2015fca,Huang:2015oca,Sun:2016nig,Shi:2017cpu}. These issues will be addressed in a forthcoming publication~\cite{in_progress}.


\acknowledgments
This work has been supported by Plan Nacional de Altas Energ\'{\i}as Spanish MINECO grants FPA2015-64041-C2-1-P,
FPA2015-64041-C2-2-P, and by Basque Government grant IT979-16. The research of E.M. is also supported by Spanish MINEICO and 
European FEDER funds grant FIS2017-85053-C2-1-P, Junta de Andaluc\'{\i}a grant FQM-225, as well as by Universidad del Pa\'{\i}s Vasco UPV/EHU 
through a Visiting Professor appointment and by Spanish MINEICO Ram\'on y Cajal Program.
M.A.V.-M. gratefully acknowledges the hospitality of the KEK Theory Center and the Department of Theoretical Physics of the University 
of the Basque Country during the early stages of this work.


\appendix

\section{The generalized transgression formula}
\label{app:transgression}

The aim of this Appendix is to summarize basic aspects of the generalized transgression formula introduced in 
Ref.~\cite{Manes:1985df}, which has been 
used at various points throughout the paper. Let us consider a family of connections $\mcA_{t}$
depending on a set of $p+2$ continuous parameters $t\equiv (t_{0},\ldots,t_{p+1})$, satisfying the constraint
\begin{align}
\sum_{r=0}^{p+1}t_{r}=1,
\end{align}
and taking values in a domain $T$. We denote by $\ell_{t}$ the substitution operator replacing the standard exterior differential $d$ by the differential in parameter space $d_{t}$
\begin{align}
\ell_{t}\equiv d_{t}{\partial\over\partial(d)},
\label{eq:ell_t_def}
\end{align}
where the odd differential operator $d_{t}$ is defined by
\begin{align}
d_{t}\equiv \sum_{r=0}^{p+1}dt_{r}{\partial\over\partial t_{r}}.
\label{eq:d_t_definition}
\end{align}
It is important to notice that the operator $\ell_{t}$ is even, since it replaces one exterior differential by another.
We consider now a generic polynomial $\mathscr{Q}$ depending on $\{\mcA_{t},\mcF_{t},d_{t}\mcA_{t},d_{t}\mcF_{t}\}$,
of degree $q$ in $d_{t}$. Then, 
the following generalized transgression formula holds~\cite{Manes:1985df}
\begin{align}
\int_{\partial T}{\ell_{t}^{p}\over p!}\mathscr{Q}=\int_{T}{\ell_{t}^{p+1}\over (p+1)!}d\mathscr{Q}+(-1)^{p+q}
d\int_{T}{\ell_{t}^{p+1}\over (p+1)!}\mathscr{Q}.
\label{eq:generalized_transgression}
\end{align}

For illustration, we apply this formula to the family of connections interpolating between $\mcA$ and $\mcB$
\begin{align}
\mcA_{t}=t\mcA+(1-t)\mcB,
\end{align} 
with $0\leq t\leq 1$. The action of the operator $\ell_{t}$ on the connection $\mcA_{t}$ and the associated field strength
$\mcF_{t}=d\mcA_{t}+\mcA_{t}^{2}$ is given by
\begin{align}
\ell_{t}\mcA_{t}&=0, \nonumber \\[0.2cm]
\ell_{t}\mcF_{t}&=d_{t}\mcA_{t}=dt\big(\mcA-\mcB\big).
\end{align}
Let us take the polynomial defined by the Chern-Simons form associated with $\mcA_{t}$
\begin{align}
\mathscr{Q}=\omega^{0}_{2n-1}(\mcA_{t}),
\end{align}
which is of degree $q=0$ in $d_{t}$. 
Since our family of connections depends on a single independent parameter ($p=0$), the generalized transgression formula \eqref{eq:generalized_transgression} in this case renders Eq. \eqref{eq:genhom1}
\begin{align}
\omega^{0}_{2n-1}(\mcA)-\omega^{0}_{2n-1}(\mcB)&
=\int_{0}^{1}\ell_{t}d\omega^{0}_{2n-1}(\mcA_{t})
+d\int_{0}^{1}\ell_{t}\omega^{0}_{2n-1}(\mcA_{t}) \nonumber \\[0.2cm]
&=\int_{0}^{1}\ell_{t}{\rm Tr\,}\mcF_{t}^{n}
+d\int_{0}^{1}\ell_{t}\omega^{0}_{2n-1}(\mcA_{t}),
\label{eq:app_DeltaOmegaellt}
\end{align}
where we have used that $d\omega^{0}_{2n-1}(\mcA_{t})={\rm Tr\,}\mcF_{t}^{n}$. 

Particularizing this expression to $\mcA=\mcA_{0}\theta+\mbA$ and $\mcB=\mbA$, we obtain Eq. \eqref{eq:genhom2}. 
To compute the first term, we just notice that 
\begin{align}
\ell_{t}d\omega^{0}_{2n-1}&=\ell_{t}{\rm Tr\,}\mcF_{t}^{n}=n{\rm Tr\,}(\mcF_{t}^{n-1}\ell_{t}\mcF_{t}) \nonumber\\[0.2cm]
&=n{\rm Tr\,}(\mcF_{t}^{n-1}d_{t}\mcA_{t})=dt\,n\,{\rm Tr\,}(\mcF_{t}^{n-1}\mcA_{0}\theta),
\end{align}
which, together with 
\begin{align}
\mcF_{t}=\mbF+t\mcA_{0}da+t(\mbD\mcA_{0})\theta,
\end{align}
leads to Eq. \eqref{eq:integral_domega0_2n-1stationary}. Here again $\mbD$ denotes the covariant derivative
with respect to the connection $\mbA$ defined in \eqref{eq:covariant_derivative_bfA}. 
As for the second term in  \eqref{eq:app_DeltaOmegaellt}, we have to remember that
$\ell_{t}$ acts on products through the Leibniz rule, replacing $\mcF_{t}$ by $d_{t}\mcA_{t}$. Thus, we can write
\begin{align}
\ell_{t}\omega^{0}_{2n-1}(\mcA_{t},\mcF_{t})&=\left.(d_{t}\mcA_{t}){\delta\over\delta\mcF}\omega^{0}_{2n-1}(\mcA,\mcF)
\right|_{\substack{\hspace*{-2.4cm}\mcA\to\mbA \\[0.1cm]\mcF\to\mbF+t\mcA_0da+t(\mbD\mcA_{0})\theta}}.
\end{align}
Taking finally into account that $d_{t}\mcA_{t}=dt\mcA_{0}\theta$, and that $\theta^{2}=0$, we retrieve 
Eq. \eqref{eq:integral_domega0_2n-1stationary_2ndterm}.


\section{Some explicit expressions}
\label{app:formulae}

To make our presentation self-contained, in this Appendix we list some relevant expressions refereed to at various points of the article.

\paragraph{A closed expression of the Bardeen counterterm.}

Taking the two-parameter family of connections  
\begin{align}
\mcA_t=t_{1}\mcA_R+t_{2}\mcA_L,
\end{align} 
a simple explicit expression for the Bardeen counterterm can be written as~\cite{Manes:1984gk}
\begin{align}
S_{2n-2}(\mcA_{R,L},\mcF_{R,L})={1\over 2}n(n-1)\int_T {\rm Str\,}\Big[(d_t\mcA_t)(d_t\mcA_t)\mcF_t^{n-2}\Big],
\label{eq:counterterm}
\end{align}
where $d_{t}$ is defined in Eq. \eqref{eq:d_t_definition} and 
the integration domain $T$ is bounded by the right triangle on the $(t_{1},t_{2})$-plane 
with vertices at $(0,0)$, $(1,0)$, and $(0,1)$. In writing this formula we have used a symmetrized trace defined
as
\begin{align}
{\rm Str\,}\Big(\mathcal{O}_{1},\ldots,\mathcal{O}_{n}\Big)={1\over n!}\sum_{\sigma\in S_{n}}{\rm sgn}(\sigma)
{\rm Tr\,}\Big(\mathcal{O}_{\sigma(1)},\ldots,\mathcal{O}_{\sigma(n)}\Big),
\end{align}
where ${\rm sgn}(\sigma)$ is the sign arising from the permutation of the forms inside the trace.

The Bardeen counterterm in four dimensions is obtained by setting $n=3$ in~\eqref{eq:counterterm}
\begin{align}
S_{4}(\mcA_{R,L},\mcF_{R,L})=3\int_{T}d^{2}t\,{\rm Str}\Big[(\mcA_{R}\mcA_{L}+\mcA_{L}\mcA_{R})\mcF_{t}\Big],
\end{align}
where
\begin{align}
\mcF_{t}=t_{1}\mcF_{R}+t_{2}\mcF_{L}+t_{1}(t_{1}-1)\mcA_{R}^{2}+t_{2}(t_{2}-1)\mcA_{L}^{2}
+t_{1}t_{2}(\mcA_{R}\mcA_{L}+\mcA_{L}\mcA_{R}).
\end{align}
After evaluating the integral, the result is
\begin{align}
S_{4}(\mcA_{R,L},\mcF_{R,L})&=\frac{1}{2}{\rm Tr}\bigg[(\mcA_L\mcA_R+\mcA_R\mcA_L)(\mcF_R+\mcF_L)  \nonumber \\[0.2cm]
&\left.+\mcA_R^3\mcA_L+ \mcA_L^3\mcA_R
+\frac{1}{2}\mcA_L\mcA_R\mcA_L\mcA_R\right].
\end{align}

\paragraph{Currents and energy-momentum tensor.}
We have seen that, upon the dimensional reduction carried out in Sec. \ref{sec:anomalousPF}, the corresponding functionals depend on the KK-invariant 
gauge fields, $\mcA_{0}$ and $\mbA$, as well as on the field strength $da$ of the Abelian KK field. Thus, upon a generic variation of the gauge field
\eqref{eq:B_general_varA}, we have the following definition of the consistent and covariant currents $\mcJ_{0}$ and $\mbJ$
\begin{align}
\delta_{B}W[\mcA_{0},\mbA,da]_{\rm inv}&=\int\limits_{S^{2n-3}}{\rm Tr\,}\Big(\mcB_{0}\mcJ_{0,\rm cov}+\mbB\mbJ_{\rm cov}\Big)+
\mbox{bulk term},
\nonumber \\[0.2cm]
\delta_{B}W[\mcA_{0},\mbA,da]_{\rm anom}&=\int\limits_{S^{2n-3}}{\rm Tr\,}\Big(\mcB_{0}\mcJ_{0,\rm cons}+\mbB\mbJ_{\rm cons}\Big).
\label{eq:general_gauge_var_Winv+anom}
\end{align}
The second equation is the variation of a local functional, already defined as an integral over the spatial manifold $S^{2n-3}$, so it
only gives boundary contributions. Notice that $\mcB_{0}$ is a zero-form, whereas $\mbB$ is a one form. Thus, the currents in Eq. \eqref{eq:general_gauge_var_Winv+anom} are respectively $(2n-3)$-forms ($\mcJ_{0,\rm cov}$ and $\mcJ_{0,\rm cons}$) and 
$(2n-4)$-forms ($\mbJ_{\rm cov}$ and $\mbJ_{\rm cons}$). The associated bona-fide currents are defined as the corresponding Hodge duals
\begin{align}
j_{0,\rm cov}&=\star \mcJ_{0,\rm cov}, \nonumber \\[0.2cm]
\boldsymbol{j}_{\rm cov}&=\star \mbJ_{\rm cov},
\end{align}
and similarly for the consistent ones. 
Notice in particular that, whereas $\boldsymbol{j}_{\rm cov}$ and $\boldsymbol{j}_{\rm cons}$ are one-forms, $j_{0,\rm cov}$ and $j_{0,\rm cons}$ are zero-forms.

We can proceed along similar lines with the energy-momentum tensor. Using Eq. \eqref{eq:metric}, we have that the mixed components 
$G_{0i}$ of the
metric are
\begin{align}
G_{0i}(\x)=-e^{2\sigma(\x)}a_{i}(\x) \hspace*{0.5cm} \Longrightarrow \hspace*{0.5cm}
{\delta\over \delta G_{0i}}=-e^{-2\sigma}{\delta\over \delta a_{i}},
\end{align}
where the second identity has to be understood as acting on functionals independent of $\sigma$. 
When applying this expression to our effective actions, we have to keep in mind that the functionals do depend on $a$
both explicitly, through $da$, and implicitly via the combination $\mbA=\boldsymbol{\mathcal{A}}-\mcA_{0}a$  [cf. Eq. \eqref{eq:boldA_form_def}].
Lowering the zero index, we find the following expression for the mixed components of the energy-momentum tensor
\begin{align}
\sqrt{-G}T_{0}^{\,\,\,i}&\equiv 
T_{0}G_{00}{\delta\over \delta G_{0i}}W[\mcA_{0},\mbA,da]=T_{0}{\delta\over \delta a_{i}}W[\mcA_{0},\boldsymbol{\mathcal{A}}-\mcA_{0}a,da]
\nonumber \\[0.2cm]
&=T_{0}\left[\left({\delta\over \delta a_{i}}\right)_{\mbA}-\mcA_{0}{\delta\over\delta A_{i}}\right]W[\mcA_{0},\mbA,da],
\label{eq:explicit_form_Ti0_fder}
\end{align}
where $G\equiv -e^{2\sigma}\det g$ is the determinant of the stationary metric \eqref{eq:metric}, 
$T_{0}$ is the equilibrium temperature, and in the
second line the functional derivative with respect to $a_{i}$ is taken at fixed $\mbA$. Applying this
to the anomalous part of the partition function, we define the $(2n-4)$-form $\mbT$ from the linear variation of $W[\mcA_{0},\mbA,da]_{\rm anom}$ 
with respect to $a$ as
\begin{align}
\delta W[\mcA_{0},\mbA,da]_{\rm anom}=\int\limits_{S^{2n-3}}\delta a\,\mbT,
\end{align}
where $\mbT$ is dual to the one form $\sqrt{-G}T_{0i}dx^{i}=e^{\sigma}\sqrt{\det g}\,g_{ij}T_{0}^{\,\,\,j}dx^{i}$.

\paragraph{Gauge transformation of the Chern-Simons form.} In order to construct the effective action for Goldstone bosons, it
is necessary to have explicit expressions for the terms of the gauge transformation of the Chern-Simons form  
given in Eq. \eqref{eq:finitetrans}. Using the homotopy 
formula given in \eqref{eq:omega0_2n-1_one_field}, it is immediate to arrive at the following equation
for the second term on the right-hand side
\begin{align}
\omega_{2n-1}^{0}(dgg^{-1},0)=(-1)^{n+1}\frac{n!(n-1)!}{(2n-1)!}{\rm Tr\,}\Big[(dgg^{-1})^{2n-1}\Big]\,.
\label{eq:winding}
\end{align} 
In the case of the third term, $\alpha_{2n-2}(\mcA,\mcF,g)$, a higher-order homotopy formula leads to~\cite{Manes:1984gk}  
\begin{align}
\alpha_{2n-2}(\mcA,\mcF,g)=n(n-1)\int_T {\rm Str\,}\Big[\mcA(dg g^{-1})\mcF_{t}^{n-2}\Big]\,,
\label{eq:alpha}
\end{align}
where here $\mcA_{t}$ denotes the two-parameter family of connections
\begin{align}
\mcA_{t}=t_{1} \mcA-t_{2} dg g^{-1},
\end{align}
and the integration domain $T$ is the same right triangle as in \eqref{eq:counterterm}.


\section{Trace identities for U(2)}
\label{app:traces}

In applying our expressions to hadronic fluids, it is useful to consider the two-flavor QCD case where the 
chiral group is U(2)$_{L}\times$U(2)$_{R}$. Here we list some relevant trace identities for the non-semisimple group U(2) 
leading to some of the expressions found in this paper. Taking the four generators 
$\{t_{0},t_{i}\}$ ($i=1,2,3$) defined in Eq. \eqref{eq:generatorsU(2)}, 
the properties of the Pauli matrices imply the identity  
\begin{align}
t_{j}t_{k}={1\over 4}\delta_{jk}\mathbb{1}+{i\over 2}\epsilon_{jk\ell}t_{\ell}.
\label{eq:tjtk_identity}
\end{align}
For arbitrary $p$- and $q$-forms $\omega_{p}$ and $\eta_{q}$ in the adjoint representation of U(2), we can immediately write
\begin{align}
{\rm Tr\,}(\omega_{p}\eta_{q})={1\over 2}({\rm Tr\,}\omega_{p})({\rm Tr\,}\eta_{q})+{\rm Tr\,}(\widehat{\omega}_{p}\widehat{\eta}_{q}),
\end{align}
where the hat indicates the components of the $p$- and $q$-forms over the SU(2) factor of U(2)$=$U(1)$\times$SU(2), namely
\begin{align}
\widehat{\omega}_{p}&\equiv \omega_{p}-{1\over 2}({\rm Tr\,}\omega_{p})\mathbb{1}.
\label{eq:hat_definition}
\end{align}

Using Eq. \eqref{eq:tjtk_identity}, we can find the corresponding identity for the trace of three $p$-, $q$-, and $r$-forms
\begin{align}
{\rm Tr\,}(\omega_{p}\eta_{q}\xi_{r})&=-{1\over 2}({\rm Tr\,}\omega_{p})({\rm Tr\,}\eta_{q})({\rm Tr\,}\xi_{r})
+{1\over 2}\Big[({\rm Tr\,}\omega_{p}){\rm Tr\,}(\eta_{q}\xi_{r})
\nonumber \\[0.2cm]
&+(-1)^{pq}({\rm Tr\,}\eta_{q}){\rm Tr\,}(\omega_{p}\xi_{r})
+{\rm Tr\,}(\omega_{p}\eta_{q})({\rm Tr\,}\xi_{r})\Big]
+{\rm Tr\,}(\widehat{\omega}_{p}\widehat{\eta}_{q}\widehat{\xi}_{r}),
\label{eq:trace3pqrforms}
\end{align}
where once again hatted quantities lie on the SU(2) factor. A similar calculation can be carried out for a trace with four
adjoint differential forms
to give
\begin{align}
{\rm Tr\,}(\omega_{p}\eta_{q}\xi_{r}\zeta_{s})&={3\over 8}({\rm Tr\,}\omega_{p})({\rm Tr\,}\eta_{q})
({\rm Tr\,}\xi_{r})({\rm Tr\,}\zeta_{s})-{3\over 8}\Big[({\rm Tr\,}\omega_{p})({\rm Tr\,}\eta_{q})
{\rm Tr\,}(\xi_{r}\zeta_{s}) \nonumber \\[0.2cm]
&+(-1)^{rs}({\rm Tr\,}\omega_{p}){\rm Tr\,}(\eta_{q}\zeta_{s})({\rm Tr\,}\xi_{r})
+({\rm Tr\,}\omega_{p}){\rm Tr\,}(\eta_{q}\xi_{r})({\rm Tr\,}\zeta_{s})
\nonumber \\[0.2cm]
&+(-1)^{pq+rs}({\rm Tr\,}\eta_{q}){\rm Tr\,}(\omega_{p}\zeta_{s})({\rm Tr\,}\xi_{r})
+(-1)^{pq}({\rm Tr\,}\eta_{q}){\rm Tr\,}(\omega_{p}\xi_{r})({\rm Tr\,}\zeta_{s})
\nonumber \\[0.2cm]
&+{\rm Tr\,}(\omega_{p}\eta_{q})({\rm Tr\,}\xi_{r})({\rm Tr\,}\zeta_{s})\Big]
+{1\over 2}\Big[({\rm Tr\,}\omega_{p}){\rm Tr\,}(\eta_{q}\xi_{r}\zeta_{s})
\label{eq:trace4pqrsforms}\\[0.2cm]
&+(-1)^{pq}({\rm Tr\,}\eta_{q}){\rm Tr\,}(\omega_{p}\xi_{r}\zeta_{s})
+(-1)^{rs}{\rm Tr\,}(\omega_{p}\eta_{q}\zeta_{s})({\rm Tr\,}\xi_{r})
\nonumber \\[0.2cm]
&+{\rm Tr\,}(\omega_{p}\eta_{q}\xi_{r})({\rm Tr\,}\zeta_{s})\Big]
+{\rm Tr\,}(\widehat{\omega}_{p}\widehat{\eta}_{q}\widehat{\xi}_{r}\widehat{\zeta}_{s}).
\nonumber
\end{align}

As a further useful example, we compute the trace 
\begin{align}
{\rm Tr\,}\omega^{5}&={\rm Tr\,}\left[\widehat{\omega}+{1\over 2}({\rm Tr\,}\omega)\mathbb{1}\right]^{5}
={\rm Tr\,}\widehat{\omega}^{5},
\end{align}
with $\omega$ a one form. To write the last equality 
we have taken into account that ${\rm Tr\,}\omega$ is an ``Abelian'' one-form and therefore
$({\rm Tr\,}\omega)^{n}=0$ for $n\geq 2$. Moreover, all terms linear in ${\rm Tr\,}\omega$ vanish, since
${\rm Tr\,}\widehat{\omega}^{4}=0$,
which follows from the cyclicity property. Finally, applying the identity
\begin{align}
{\rm Tr\,}(t_{j}t_{k}t_{\ell}t_{m}t_{n})={i\over 16}\Big(\delta_{jk}\epsilon_{\ell mn}
+\delta_{\ell m}\epsilon_{jkn}+\delta_{kn}\epsilon_{j\ell m}+\delta_{jn}\epsilon_{k\ell m}\Big),
\label{eq:trace5lambdas}
\end{align}
we see that, given that $\omega$ is a one-form,
\begin{align}
{\rm Tr\,}\omega^{5}=0.
\label{eq:tromega5=0}
\end{align}

Finally, we evaluate the traces ${\rm Tr\,}(u\omega^{3})$ and ${\rm Tr\,}(u\omega^{4})$, with $u$ and $\omega$
zero- and one-forms respectively. In the first case
\begin{align}
{\rm Tr\,}(u\,\omega^{3})&={\rm Tr\,}(u\,\widehat{\omega}^{3})+{1\over 2}{\rm Tr\,}(u\,\widehat{\omega}^{2}){\rm Tr\,}\omega,
\end{align}
whereas using 
\begin{align}
{\rm Tr\,}(t_{i}t_{j}t_{k}t_{\ell})&={1\over 8}\Big(
\delta_{ij}\delta_{k\ell}-\delta_{ik}\delta_{j\ell}+\delta_{i\ell}\delta_{jk}
\Big),
\label{eq:trace4lambdas}
\end{align}
we find that 
\begin{align}
{\rm Tr\,}(u\,\omega^{3})&={1\over 2}({\rm Tr\,}u){\rm Tr\,}\widehat{\omega}^{3}+
{1\over 2}{\rm Tr\,}(u\,\widehat{\omega}^{2}){\rm Tr\,}\omega.
\label{eq:truomega3_pre}
\end{align}
As a final step, we can express the right-hand side of this equation in terms of hatless quantities. Using Eq. \eqref{eq:trace3pqrforms},
we find ${\rm Tr\,}\widehat{\omega}^{3}={\rm Tr\,}\omega^{3}$.
In addition, the presence of ${\rm Tr\,}\omega$ in the second term of the right-hand side of \eqref{eq:truomega3_pre} allows to
remove the hat on the prefactor, to 
finally write
\begin{align}
{\rm Tr\,}(u\,\omega^{3})&={1\over 2}({\rm Tr\,}u){\rm Tr\,}\omega^{3}+
{1\over 2}{\rm Tr\,}(u\,\omega^{2}){\rm Tr\,}\omega.
\label{eq:truomega3}
\end{align}
A similar computation leads to
\begin{align}
{\rm Tr\,}(u\,\omega^{4})&=0.
\end{align}


\bibliographystyle{JHEP}
\bibliography{NAHF_v2}

\providecommand{\href}[2]{#2}\begingroup\raggedright\begin{thebibliography}{10}

\bibitem{Zumino:1983ew}
B.~Zumino, {\it {Chiral Anomalies and Differential Geometry}},  in {\em
  {``Relativity, groups and topology''}}, {Elsevier}, 1983.

\bibitem{AlvarezGaume:1985ex}
L.~\'Alvarez-Gaum\'e, {\it {An Introduction to Anomalies}},  in {\em
  {``Fundamental Problems of Gauge Field Theory''}}, {Plenum Press}, 1985.

\bibitem{Bertlmann:1996xk}
R.~A. Bertlmann, {\em {Anomalies in Quantum Field Theory}}.
\newblock Oxford University Press, 1996.

\bibitem{Fujikawa:2004cx}
K.~Fujikawa and H.~Suzuki, {\em {Path Integrals and Quantum Anomalies}}.
\newblock Oxford University Press, 2004.

\bibitem{Harvey:2005it}
J.~A. Harvey, {\it {TASI Lectures on Anomalies}},
\newblock \href{http://xxx.lanl.gov/abs/hep-th/0509097}{{\tt hep-th/0509097}}.

\bibitem{Son:2009tf}
D.~T. Son and P.~Surowka, {\it {Hydrodynamics with Triangle Anomalies}},  {\em
  Phys. Rev. Lett.} {\bf 103} (2009) 191601,
  [\href{http://xxx.lanl.gov/abs/0906.5044}{{\tt arXiv:0906.5044}}].

\bibitem{Neiman:2010zi}
Y.~Neiman and Y.~Oz, {\it {Relativistic Hydrodynamics with General Anomalous
  Charges}},  {\em JHEP} {\bf 03} (2011) 023,
  [\href{http://xxx.lanl.gov/abs/1011.5107}{{\tt arXiv:1011.5107}}].

\bibitem{Sadofyev:2010pr}
A.~V. Sadofyev and M.~V. Isachenkov, {\it {The Chiral magnetic effect in
  hydrodynamical approach}},  {\em Phys. Lett.} {\bf B697} (2011) 404,
  [\href{http://xxx.lanl.gov/abs/1010.1550}{{\tt arXiv:1010.1550}}].

\bibitem{Kirilin:2012mw}
V.~P. Kirilin, A.~V. Sadofyev, and V.~I. Zakharov, {\it {Chiral Vortical Effect
  in Superfluid}},  {\em Phys. Rev.} {\bf D86} (2012) 025021,
  [\href{http://xxx.lanl.gov/abs/1203.6312}{{\tt arXiv:1203.6312}}].

\bibitem{Fukushima:2012vr}
K.~Fukushima, {\it {Views of the Chiral Magnetic Effect}},  in {\em {``Strongly
  Interacting Matter in Magnetic Fields''}} ({D.~Kharzeev, K.~Landsteiner,
  A.~Schmitt, and H.-U.~Yee}, ed.), {Springer Verlag}, 2013.
\newblock \href{http://xxx.lanl.gov/abs/1209.5064}{{\tt arXiv:1209.5064}}.

\bibitem{Zakharov:2012vv}
V.~I. Zakharov, {\it {Chiral Magnetic Effect in Hydrodynamic Approximation}},
  in {\em {``Strongly Interacting Matter in Magnetic Fields''}} ({D.~Kharzeev,
  K.~Landsteiner, A.~Schmitt, and H.-U.~Yee}, ed.), {Springer Verlag}, 2013.
\newblock \href{http://xxx.lanl.gov/abs/1210.2186}{{\tt arXiv:1210.2186}}.

\bibitem{Kharzeev:2015znc}
D.~E. Kharzeev, J.~Liao, S.~A. Voloshin, and G.~Wang, {\it {Chiral magnetic and
  vortical effects in high-energy nuclear collisions - A status report}},  {\em
  Prog. Part. Nucl. Phys.} {\bf 88} (2016) 1,
  [\href{http://xxx.lanl.gov/abs/1511.04050}{{\tt arXiv:1511.04050}}].

\bibitem{Landsteiner:2016led}
K.~Landsteiner, {\it {Notes on Anomaly Induced Transport}},  {\em Acta Phys.
  Polon.} {\bf B47} (2016) 2617,
  [\href{http://xxx.lanl.gov/abs/1610.04413}{{\tt arXiv:1610.04413}}].

\bibitem{Gooth:2017mbd}
J.~Gooth {\em et.~al.}, {\it {Experimental signatures of the mixed
  axial-gravitational anomaly in the Weyl semimetal NbP}},  {\em Nature} {\bf
  547} (2017) 324, [\href{http://xxx.lanl.gov/abs/1703.10682}{{\tt
  arXiv:1703.10682}}].

\bibitem{Banerjee:2012iz}
N.~Banerjee, J.~Bhattacharya, S.~Bhattacharyya, S.~Jain, S.~Minwalla, and
  T.~Sharma, {\it {Constraints on Fluid Dynamics from Equilibrium Partition
  Functions}},  {\em JHEP} {\bf 09} (2012) 046,
  [\href{http://xxx.lanl.gov/abs/1203.3544}{{\tt arXiv:1203.3544}}].

\bibitem{Jensen:2012jh}
K.~Jensen, M.~Kaminski, P.~Kovtun, R.~Meyer, A.~Ritz, and A.~Yarom, {\it
  {Towards hydrodynamics without an entropy current}},  {\em Phys. Rev. Lett.}
  {\bf 109} (2012) 101601, [\href{http://xxx.lanl.gov/abs/1203.3556}{{\tt
  arXiv:1203.3556}}].

\bibitem{Jensen:2012kj}
K.~Jensen, R.~Loganayagam, and A.~Yarom, {\it {Thermodynamics, gravitational
  anomalies and cones}},  {\em JHEP} {\bf 02} (2013) 088,
  [\href{http://xxx.lanl.gov/abs/1207.5824}{{\tt arXiv:1207.5824}}].

\bibitem{Jensen:2013kka}
K.~Jensen, R.~Loganayagam, and A.~Yarom, {\it {Anomaly inflow and thermal
  equilibrium}},  {\em JHEP} {\bf 05} (2014) 134,
  [\href{http://xxx.lanl.gov/abs/1310.7024}{{\tt arXiv:1310.7024}}].

\bibitem{Wess:1971yu}
J.~Wess and B.~Zumino, {\it {Consequences of anomalous Ward identities}},  {\em
  Phys. Lett.} {\bf 37B} (1971) 95.

\bibitem{Witten:1983tw}
E.~Witten, {\it {Global Aspects of Current Algebra}},  {\em Nucl. Phys.} {\bf
  B223} (1983) 422.

\bibitem{Manes:1984gk}
J.~L. Ma\~nes, {\it {Differential Geometric Construction of the Gauged
  {Wess-Zumino} Action}},  {\em Nucl. Phys.} {\bf B250} (1985) 369.

\bibitem{Haehl:2013hoa}
F.~M. Haehl, R.~Loganayagam, and M.~Rangamani, {\it {Effective actions for
  anomalous hydrodynamics}},  {\em JHEP} {\bf 03} (2014) 034,
  [\href{http://xxx.lanl.gov/abs/1312.0610}{{\tt arXiv:1312.0610}}].

\bibitem{Lin:2011mr}
S.~Lin, {\it {On the anomalous superfluid hydrodynamics}},  {\em Nucl. Phys.}
  {\bf A873} (2012) 28--46, [\href{http://xxx.lanl.gov/abs/1104.5245}{{\tt
  arXiv:1104.5245}}].

\bibitem{Nair:2011mk}
V.~P. Nair, R.~Ray, and S.~Roy, {\it {Fluids, Anomalies and the Chiral Magnetic
  Effect: A Group-Theoretic Formulation}},  {\em Phys. Rev.} {\bf D86} (2012)
  025012, [\href{http://xxx.lanl.gov/abs/1112.4022}{{\tt arXiv:1112.4022}}].

\bibitem{Bhattacharyya:2012xi}
S.~Bhattacharyya, S.~Jain, S.~Minwalla, and T.~Sharma, {\it {Constraints on
  Superfluid Hydrodynamics from Equilibrium Partition Functions}},  {\em JHEP}
  {\bf 01} (2013) 040, [\href{http://xxx.lanl.gov/abs/1206.6106}{{\tt
  arXiv:1206.6106}}].

\bibitem{Lublinsky:2009wr}
M.~Lublinsky and I.~Zahed, {\it {Anomalous Chiral Superfluidity}},  {\em Phys.
  Lett.} {\bf B684} (2010) 119--122,
  [\href{http://xxx.lanl.gov/abs/0910.1373}{{\tt arXiv:0910.1373}}].

\bibitem{Neiman:2011mj}
Y.~Neiman and Y.~Oz, {\it {Anomalies in Superfluids and a Chiral Electric
  Effect}},  {\em JHEP} {\bf 09} (2011) 011,
  [\href{http://xxx.lanl.gov/abs/1106.3576}{{\tt arXiv:1106.3576}}].

\bibitem{Lin:2011aa}
S.~Lin, {\it {An anomalous hydrodynamics for chiral superfluid}},  {\em Phys.
  Rev.} {\bf D85} (2012) 045015, [\href{http://xxx.lanl.gov/abs/1112.3215}{{\tt
  arXiv:1112.3215}}].

\bibitem{Hoyos:2014nua}
C.~Hoyos, B.~S. Kim, and Y.~Oz, {\it {Odd Parity Transport in Non-Abelian
  Superfluids from Symmetry Locking}},  {\em JHEP} {\bf 10} (2014) 127,
  [\href{http://xxx.lanl.gov/abs/1404.7507}{{\tt arXiv:1404.7507}}].

\bibitem{Jain:2016rlz}
A.~Jain, {\it {Theory of non-Abelian superfluid dynamics}},  {\em Phys. Rev.}
  {\bf D95} (2017) 121701, [\href{http://xxx.lanl.gov/abs/1610.05797}{{\tt
  arXiv:1610.05797}}].

\bibitem{in_progress}
J.~L. Ma\~nes, E.~Meg\'{\i}as, M.~Valle, and M.~{\'A}. V\'azquez-Mozo, {\em
  Anomalous Currents of Nuclear Matter in the Chiral Limit}.
\newblock To appear.

\bibitem{Manes:1985df}
J.~Ma\~nes, R.~Stora, and B.~Zumino, {\it {Algebraic Study of Chiral
  Anomalies}},  {\em Commun. Math. Phys.} {\bf 102} (1985) 157.

\bibitem{Nakahara:2003nw}
M.~Nakahara, {\em {Geometry, Topology and Physics (2nd edition)}}.
\newblock Taylor \& Francis, 2003.

\bibitem{AlvarezGaume:1983cs}
L.~\'Alvarez-Gaum\'e and P.~Ginsparg, {\it {The Topological Meaning of
  Nonabelian Anomalies}},  {\em Nucl. Phys.} {\bf B243} (1984) 449.

\bibitem{Bardeen:1969md}
W.~A. Bardeen, {\it {Anomalous Ward identities in spinor field theories}},
  {\em Phys. Rev.} {\bf 184} (1969) 1848.

\bibitem{Kaiser:2000ck}
R.~Kaiser, {\it {Anomalies and WZW term of two flavor QCD}},  {\em Phys. Rev.}
  {\bf D63} (2001) 076010, [\href{http://xxx.lanl.gov/abs/hep-ph/0011377}{{\tt
  hep-ph/0011377}}].

\bibitem{Bardeen:1984pm}
W.~A. Bardeen and B.~Zumino, {\it {Consistent and Covariant Anomalies in Gauge
  and Gravitational Theories}},  {\em Nucl. Phys.} {\bf B244} (1984) 421--453.

\bibitem{Callan:1984sa}
C.~G. Callan, Jr. and J.~A. Harvey, {\it {Anomalies and Fermion Zero Modes on
  Strings and Domain Walls}},  {\em Nucl. Phys.} {\bf B250} (1985) 427.

\bibitem{Haber:1981ts}
H.~E. Haber and H.~A. Weldon, {\it {Finite Temperature Symmetry Breaking as
  Bose-Einstein Condensation}},  {\em Phys. Rev.} {\bf D25} (1982) 502.

\bibitem{Yamada:2006rx}
D.~Yamada and L.~G. Yaffe, {\it {Phase diagram of N=4 super-Yang-Mills theory
  with R-symmetry chemical potentials}},  {\em JHEP} {\bf 09} (2006) 027,
  [\href{http://xxx.lanl.gov/abs/hep-th/0602074}{{\tt hep-th/0602074}}].

\bibitem{Fukushima:2008xe}
K.~Fukushima, D.~E. Kharzeev, and H.~J. Warringa, {\it {The Chiral Magnetic
  Effect}},  {\em Phys. Rev.} {\bf D78} (2008) 074033,
  [\href{http://xxx.lanl.gov/abs/0808.3382}{{\tt arXiv:0808.3382}}].

\bibitem{Landsteiner:2012kd}
K.~Landsteiner, E.~Meg\'{\i}as, and F.~Pe\~na Ben\'{\i}tez, {\it {Anomalous
  Transport from Kubo Formulae}},  in {\em {``Strongly Interacting Matter in
  Magnetic Fields''}} ({D.~Kharzeev, K.~Landsteiner, A.~Schmitt, and
  H.-U.~Yee}, ed.), {Springer Verlag}, 2013.
\newblock \href{http://xxx.lanl.gov/abs/1207.5808}{{\tt arXiv:1207.5808}}.

\bibitem{Kawai:1984mx}
H.~Kawai and S.~H.~H. Tye, {\it {Chiral Anomalies, Effective Lagrangian and
  Differential Geometry}},  {\em Phys. Lett.} {\bf 140B} (1984) 403.

\bibitem{Fukushima:2012fg}
K.~Fukushima and K.~Mameda, {\it {Wess-Zumino-Witten action and photons from
  the Chiral Magnetic Effect}},  {\em Phys. Rev.} {\bf D86} (2012) 071501,
  [\href{http://xxx.lanl.gov/abs/1206.3128}{{\tt arXiv:1206.3128}}].

\bibitem{Son:2007ny}
D.~T. Son and M.~A. Stephanov, {\it {Axial anomaly and magnetism of nuclear and
  quark matter}},  {\em Phys. Rev.} {\bf D77} (2008) 014021,
  [\href{http://xxx.lanl.gov/abs/0710.1084}{{\tt arXiv:0710.1084}}].

\bibitem{Andrianov:2017ely}
A.~Andrianov, V.~Andrianov, and D.~Espriu, {\it {Chiral imbalance in QCD}},
  {\em EPJ Web Conf.} {\bf 138} (2017) 01007.

\bibitem{AlvarezGaume:1983ig}
L.~\'Alvarez-Gaum\'e and E.~Witten, {\it {Gravitational Anomalies}},  {\em
  Nucl. Phys.} {\bf B234} (1984) 269.

\bibitem{AlvarezGaume:1984dr}
L.~\'Alvarez-Gaum\'e and P.~Ginsparg, {\it {The Structure of Gauge and
  Gravitational Anomalies}},  {\em Annals Phys.} {\bf 161} (1985) 423.

\bibitem{Yin:2015fca}
Y.~Yin and J.~Liao, {\it {Hydrodynamics with chiral anomaly and charge
  separation in relativistic heavy ion collisions}},  {\em Phys. Lett.} {\bf
  B756} (2016) 42--46, [\href{http://xxx.lanl.gov/abs/1504.06906}{{\tt
  arXiv:1504.06906}}].

\bibitem{Huang:2015oca}
X.-G. Huang, {\it {Electromagnetic fields and anomalous transports in heavy-ion
  collisions --- A pedagogical review}},  {\em Rept. Prog. Phys.} {\bf 79}
  (2016) 076302, [\href{http://xxx.lanl.gov/abs/1509.04073}{{\tt
  arXiv:1509.04073}}].

\bibitem{Sun:2016nig}
Y.~Sun, C.~M. Ko, and F.~Li, {\it {Anomalous transport model study of chiral
  magnetic effects in heavy ion collisions}},  {\em Phys. Rev.} {\bf C94}
  (2016) 045204, [\href{http://xxx.lanl.gov/abs/1606.05627}{{\tt
  arXiv:1606.05627}}].

\bibitem{Shi:2017cpu}
S.~Shi, Y.~Jiang, E.~Lilleskov, and J.~Liao, {\it {Anomalous Chiral Transport
  in Heavy Ion Collisions from Anomalous-Viscous Fluid Dynamics}},  {\em Annals
  Phys.} {\bf 394} (2018) 50, [\href{http://xxx.lanl.gov/abs/1711.02496}{{\tt
  arXiv:1711.02496}}].

\end{thebibliography}\endgroup

\end{document}